\documentclass[12pt]{article}
\usepackage[T1]{fontenc}
\usepackage{amssymb}
\usepackage{slashed}
\usepackage{cancel}
\usepackage{mathrsfs}
\usepackage{stmaryrd}
\usepackage{mathtools} 
\usepackage[bbgreekl]{mathbbol}

\newcommand{\orcid}[1]{\href{https://orcid.org/#1}{\includegraphics[width=8pt]{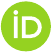}}}

\usepackage{amsthm}
\theoremstyle{definition}

\theoremstyle{plain}

\usepackage{cancel}
\usepackage{extarrows}

\newcommand{\pth}{\text{\normalfont\TH\/}}

\usepackage[colorlinks=true,linkcolor=blue,citecolor=blue,linktocpage=true]{hyperref}
\usepackage[numbers,sort&compress]{natbib}
\usepackage{tocloft}
\usepackage{titlesec}

\usepackage{authblk}
\usepackage{blindtext}

\usepackage{geometry}
\usepackage{physics}

\newcommand{\pa}{\mathop{}\!\partial}

\newcommand{\R}{\mathbb{R}}

\newcommand{\Diff}{\text{Diff}}

\def\be#1\ee{\begin{align}#1\end{align}}

\def\heq{\,\hat{=}\,}

\newcommand\nueq{\stackrel{\mathclap{\normalfont\tiny\mbox{$\scr{H}$}}}{=}}

\newcommand{\vardelta}{\bbdelta}

\newcommand{\mr}{\mathring}

\newcommand{\bd}{\boldsymbol}

\newcommand{\dext}{\text{d}}

\newcommand{\cM}{\mathcal{M}}
\newcommand{\cN}{\mathcal{N}}

\newcommand{\cA}{\mathcal{A}}

\newcommand{\cO}{\mathcal{O}}
\newcommand{\Lie}{\mathcal{L}}
\newcommand{\cW}{\mathcal{W}}
\newcommand{\cE}{\mathcal{E}}

\newcommand{\cQ}{\mathcal{Q}}
\newcommand{\cT}{\mathcal{T}}
\newcommand{\cP}{\mathcal{P}}
\newcommand{\cJ}{\mathcal{J}}

\newcommand{\scr}{\mathscr}

\newcommand{\eps}{{\bd{\epsilon}}}

\newcommand{\f}{\frac}

\def\CFL{\mathrm{CFL}}
\def\RZ{\mathrm{RZ}}

\def\scri{\mathscr{I}}

\newcommand{\mtn}{\mr{\theta}^{(n)}}
\newcommand{\mpi}{\mr{\pi}}

\newcommand{\thell}{\theta^{(\ell)}}
\newcommand{\mthell}{\mr{\theta}^{(\ell)}}

\usepackage{tikz-cd}

\newgeometry{left = 3cm, right=3cm}

\allowdisplaybreaks

\begin{document}

\title{\Large{\textbf{Higher spin dynamics in Weyl reference frame}}}

\author[ ]{\ \ \quad Gianfranco De Simone \orcid{0009-0004-5342-7670}
\ \ \quad
\footnote{\href{mailto:gianfranco.desimone@uniud.it}{gianfranco.desimone@uniud.it}
}
}

\date{}
\affil[ ]{\small\emph{Università degli Studi di Udine, via Palladio 8, I-33100 Udine, Italy}}
\affil[ ]{\small\emph{National Institute for Nuclear Physics (INFN), Sezione di Trieste, Via Valerio 2, 34127, Italy}}

\maketitle

\begin{abstract}
In this manuscript we derive the $w_{1+\infty}$ algebra on an arbitrary null hypersurface located at finite distance. A straightforward integration of the evolution Bianchi identities is performed in a suitable dynamical reference frame, dubbed the Weyl reference frame. In this reference frame, the extraction of the (linear) higher spin charge bracket is similar to the asymptotic null case, except that the corner metric evolves non trivially in time. In other words, boundary degrees of freedom that are frozen at null infinity due to the boundary conditions now become part of the dynamics. This feature is reflected in the appearance of non-local terms in the brackets. In order to avoid the presence of these non-local terms in the charge bracket, we absorb this non-locality - which encodes nothing other than the history of the corner metric - into the definition of the higher spin charges. Moreover, a new perspective on the memory effect at finite distance is suggested. In conclusion, we believe that this treatment can also be applied to asymptotically (A)dS$_{4}$ spacetimes, where the presence of a non-vanishing cosmological constant leads to the introduction of a cosmological reference frame.
\end{abstract}

\thispagestyle{empty}
\newpage
\hfill
\tableofcontents
\thispagestyle{empty}

\setcounter{page}{1}
\section*{Introduction}\addcontentsline{toc}{section}{Introduction}
Symmetries represent one of the central pillars of theoretical physics. A complete knowledge of the symmetry content of a theory allows one to label physical states and classify field configurations according to the irreducible representations of the underlying symmetry group. Furthermore, symmetries dictate the structure of interactions, constrain the form of allowed operators, and give rise to conserved charges through Noether’s theorem, thereby defining the observable content of the physical system. Therefore, a comprehensive characterization of these symmetries is crucial prior to quantization. As is well known, one of the most challenging problems in fundamental physics is the quantization of gravity; hence, unraveling the symmetries of gravitational theories serves as a necessary first step toward achieving this goal. In recent years, null hypersurfaces have emerged as a prominent arena for studying these gravitational symmetries. At finite distances, analyzing the canonical phase space on null surfaces has revealed rich, non-trivial corner symmetry algebras that govern local gravitational degrees of freedom and boundary dynamics \cite{Chandrasekaran:2018aop, Chandrasekaran:2021hxc, Ciambelli:2023mir, Ciambelli:2024swv, DeSimone:2026nig}. Taking the asymptotic limit of these structures toward future null infinity $\scri^+$ bridges local surface geometry with the program of celestial holography \cite{Strominger:2017zoo, Raclariu:2021zjz, Pasterski:2021rjz}. In this framework, four-dimensional scattering amplitudes in asymptotically flat spacetimes are mapped via Mellin transforms onto correlation functions of a two-dimensional celestial conformal field theory living on the celestial sphere \cite{Pasterski:2016qvg, Pasterski:2017kqt, Cheung:2016iub}. Crucially, the soft graviton theorems of quantum field theory translate into Ward identities for an infinite hierarchy of asymptotic symmetries \cite{Strominger:2013jfa, He:2014laa}. Beyond the familiar supertranslations and superrotations of the extended BMS group \cite{Bondi:1962px, Sachs:1962wk, Barnich:2009se}, organizing the positive-helicity soft graviton modes into conformal primary currents reveals an underlying chiral symmetry \cite{Guevara:2021abz,  Strominger:2021mtt}. This infinite tower of symmetries generates the wedge algebra of $w_{1+\infty}$, which acts as a master organizing principle for hard and soft interactions in flat-space gravity \cite{Freidel:2021ytz, Geiller:2024bgf, Adamo:2021lrv, Kmec:2024nmu, Cresto:2024fhd, Cresto:2024mne}.

In particular, the authors in \cite{Ruzziconi:2025fuy}  identify the $w_{1+\infty}$ algebra in the subleading phase space of an arbitrary null hypersurface by imposing self-duality conditions designed to recover the Ashtekar–Streubel symplectic form \cite{Ashtekar:1981bq} at finite distance. As emphasized in \cite{Ruzziconi:2025fuy, DeSimone:2026nig}, however, there are substantial differences between the finite-distance and asymptotic cases. Foremost among these is the presence at finite distance of genuine dynamical degrees of freedom that are absent asymptotically as a consequence of the boundary conditions imposed at $\scri^+$. These degrees of freedom complicate the structure of the Bianchi identities, entering both the evolution operator of the Weyl scalars and the covariant derivative on the corner. A related distinction, noted in [14], concerns the memory effect: at null infinity the memory effect appears only at subleading order, whereas on a black hole horizon $\mathscr{H}^+$ it affects the leading order of the horizon structure itself. Although a black hole memory tensor was defined in \cite{Rahman:2019bmk}, its relation to near-horizon supertranslations — and to the associated charges and fluxes — has not yet been established.

In summary, the complications that arise at finite distance, or more generally on an arbitrary null hypersurface, stem from the presence of new boundary degrees of freedom associated with the non-trivial time evolution of the boundary metric. In our view, these unfrozen degrees of freedom provide the natural building blocks for a dynamical reference frame. Indeed, such a frame — the dressing time reference frame — was already constructed from these degrees of freedom in \cite{Ciambelli:2023mir, Ciambelli:2024swv, Freidel:2025ous}. Dynamical reference frames are not ordinary coordinate systems but are instead built out of the physical degrees of freedom already present in the theory \cite{Goeller:2022rsx}; that is, they represent non-fixed, field-dependent coordinate systems.

In this work, we describe dynamical reference frames from the perspective of embedding fields, following the construction of \cite{Donnelly:2016auv, Speranza:2017gxd, Freidel:2021dxw, Ciambelli:2021nmv,  Carrozza:2021gju, Carrozza:2022xut}. We then introduce a field-dependent diffeomorphism defining a dynamical reference frame that we call the \emph{Weyl reference frame} (WRF). Under suitable assumptions, the bulk symplectic form can be written in a form that closely resembles the Ashtekar–Streubel 2-form at null infinity, with the role of the news tensor played by the longitudinal shear. As we argue, describing the system through dynamical reference frames naturally produces a corner symplectic form, defining an extended phase space. While in this work we restrict attention to the bulk term of the symplectic 2-form, the corner term plays a fundamental role in the full description of the system, since it encodes the information exchanged between the physical and reference manifolds — that is, the edge modes.

The Weyl reference frame, as constructed here, has the property that the Bianchi identities take the form
\begin{equation}
    \pa_{\tilde{v}} \tilde{\cQ}_{s} = \tilde{\eth}'_{\scr{C}} \tilde{\cQ}_{s+1} - (s+1)\tilde{\lambda}\tilde{\cQ}_{s+2},
\end{equation}
where $\tilde v$ is the dressing time, $\eth'_{\scr{C}}$ is the dressed Weyl covariant derivative, $\tilde\lambda$ is the dressed transversal shear, and $\tilde{\cQ}_s$, for $s = -1, 0, 1, 2$, are the dressed charges in the WRF. To recover the same structure found at null infinity — and thereby apply the same computational machinery — non-local terms must be added to the dressed charges $\tilde{\cQ}_s$. These terms originate from the non-commutativity of the Weyl covariant derivative with the (inverse) time-derivative operator, a non-commutativity that encodes the history of the boundary metric. The corresponding non-local corrections to the spin-$s$ dressed charges obey the recursive relation
\begin{equation}
\begin{aligned}
\tilde{\cM}_s &= [\pa_{\tilde{v}}^{-1}, \tilde{\eth}'^{2+s}_{\scr{C}}] \pa_{\tilde{v}}^{-s}\tilde{\sigma} + \pa_{\tilde{v}}^{-1}\tilde{\eth}'_{\scr{C}}\tilde{\cM}_{s-1} - (s+1)\pa_{\tilde{v}}^{-1}(\tilde{\lambda}\tilde{\cM}_{s-2})\\
&\qquad - [\pa_{\tilde{v}}^{-1}, \tilde{\eth}'_{\scr{C}}] \pa_{\tilde{v}}^{-1} (\tilde{\lambda}\tilde{\cQ}_{s-3}),
\end{aligned}
\end{equation}
with $\tilde{\cM}_{-1} = [\pa_{\tilde{v}}^{-1},\tilde{\eth}'_{\scr{C}}]\tilde{\cN}$ and $\tilde{\cM}_{-2}=0$. Apart from these differences, the computation of the $w_{1+\infty}$ algebra at finite distance proceeds by the same methodology employed in \cite{Freidel:2021ytz, Geiller:2024bgf} , yielding the bracket, truncated at linear level,
\begin{equation}
\{\tilde{Q}_{s_1}(\tilde{\tau}_1), \tilde{Q}_{s_2}(\tilde{\tau}_2)\}^1_{\tilde{\mu}_{\infty}} = (s_2+1)\tilde{Q}_{s_2+s_1-1}(\tilde{\tau}_2 \tilde{\eth}'_{\scr{C}} \tilde{\tau}_1) -(s_1+1)\tilde{Q}_{s_2+s_1-1}(\tilde{\tau}_1 \tilde{\eth}'_{\scr{C}}\tilde{\tau}_2),
\end{equation}
where $\{\cdot, \cdot\}_{\tilde{\mu}_\infty}=\tilde{\mu}_\infty^{-1} \{\cdot, \cdot\}$, with $\tilde{\mu}_{\infty}$ the value of the transversal expansion as $\tilde{v}\to+\infty$.

\vspace{0.2cm}

The paper is organized as follows. Section \ref{1} reviews the fundamental concepts of null geometry, summarizing the main results of \cite{DeSimone:2026nig} and developing the symplectic form in the spin-coefficient formalism. The Weyl and boost transformations, together with the associated covariant operators, are also discussed, and the section closes with a review of the dressing-time proposals available in the literature, which motivates the construction of Section \ref{sec2}.

Section \ref{sec2} opens with the description of dynamical reference frames from the embedding field perspective. Building on this, we introduce the Weyl reference frame and its associated dressing time, write the symplectic form in the WRF, and construct the corresponding non-local charges.

In Section \ref{sec3} we apply the methodology of \cite{Freidel:2021dfs, Freidel:2021ytz, Geiller:2024bgf} to compute the action of the non-local charges on the gravitational modes. Once a general expression for the renormalized spin-$s$ charges is obtained, the linear-level Poisson bracket between two higher-spin charges follows as a direct generalization of the asymptotic $w_{1+\infty}$ algebra, with the covariant derivative on the celestial sphere replaced by the Weyl covariant derivative. Although the computations closely parallel those of \cite{Freidel:2021ytz}, we present them in full up to the point where time dependence enters, since care must be taken with the relative ordering of the time-derivative and covariant-derivative operators.

\vspace{0.3cm}

\noindent
\emph{Notation and conventions}: We use the mostly-plus metric signature $(-+++)$ and set $c = 1$, $\varkappa = \sqrt{32\pi G}$. Four-dimensional spacetime indices are denoted by Greek letters $\mu, \nu, \sigma, \dots$; indices on the null horizon by lowercase Latin letters from the middle of the alphabet, $i, j, k, \dots$; and indices on the spacelike codimension-2 surface $\scr{S}$ by lowercase Latin letters from the beginning of the alphabet, $a, b, c, \dots$. Uppercase Latin letters $A, B, C, \dots$ denote spinor indices. Quantities defined in the Weyl reference frame are denoted with a tilde. The on-shell symbol is denoted by $\heq$, and equalities holding on the horizon ($\rho = 0$) are denoted by $\overset{\mathscr H}{=}$. Quantities evaluated on the horizon are marked with a circle, $\mr{a}\nueq a$; in the WRF, since every quantity is understood to be evaluated on the horizon, we drop the circle and write $\mr{\tilde{a}}= \tilde{a}$. The local coordinate system used on $\scr{M}$ is ($v,\rho,z,\bar{z}$) and we mostly write $z=(z, \bar{z})$.

\section{Geometry and symmetry of null surfaces}\label{1}
To begin our treatment, let us first recall the fundamental ingredients of null geometries \cite{Gourgoulhon:2005ng}. Consider a three dimensional null hyper-surface $\scr{H}$ embedded into a four dimensional spacetime manifold $(\scr{M}, g_{\mu\nu}, \nabla_\mu)$ via the embedding map
\begin{equation}
    \Pi:\scr{H}\to\scr{M}.
\end{equation}
Assuming the topology of the null hypersurface to be $\scr{H}\simeq \scr{S}\times \R$, where $\scr{S}$ is a compact codimension-2 surface (i.e., the corner), the induced metric on the corner is denoted by $q_{ab}$. The normal space to $\scr{S}$ is spanned by two future-directed null vectors $\ell$ and $n$, with $\ell\cdot n=-1$, defined up to a local rescaling
\begin{equation}
\ell\to e^{\lambda_L}\ell \qquad\text{and} \qquad n\to e^{-\lambda_L}n,
\label{resc_ln}
\end{equation}
where $\lambda_L(x)$ is a smooth function. In particular, the pair $(q_{ab}, \ell)$ defines a Carrollian structure and the vector $\ell$ is called the Carrollian vector.
Given the null vectors normal to $\scr{S}$, one can define the relative extrinsic curvatures
\begin{equation}
    K^{(k)}_{ab} = \sigma^{(k)}_{ab} + \f{1}{2}q_{ab}\theta^{(k)},
\end{equation}
where $k$ can be $\ell$ or $n$. Then, one introduces the rotational 1-form $\varpi_\mu$ and the non-affinity parameter $\kappa_{(\ell)}$, defined as follows
\begin{equation}
\varpi_\mu = - n_\nu \nabla_\mu \ell^\nu \qquad \text{and} \qquad    \ell^\mu\nabla_\mu \ell^\nu = \kappa_{(\ell)}\ell^\nu,
\end{equation}
respectively, with $\ell^\mu\varpi_\mu = \kappa_{(\ell)}$. In particular, the spatial projection of the rotational 1-form, i.e., $\pi_a = q^{\ \mu}_{a} \varpi_\mu$, is called the Hajicek field. These are the fundamental geometric quantities to analyse null hypersurfaces. \\
Now, since we are interested in the near-horizon geometry, taking $n$ the null vector pointing in the bulk of the spacetime, the behaviour of the null tetrad in this region is \cite{DeSimone:2026nig},
\begin{equation}
n =\pa_\rho, \qquad \ell = \pa_v + o(1),\qquad m =  \mr{m}^a \pa_a+ o(1),\qquad \bar{m} =  \mr{\bar{m}}^a \pa_a+ o(1),
\end{equation}
and by employing the Newman-Penrose formalism, the previously defined geometric quantities are now encoded into the spin coefficients, whose behaviour on $\scr{H}$ is
\begin{equation}
\begin{aligned}
    \kappa &=  O(\rho), \qquad& \varrho &= -\f{1}{2}\mthell + O(\rho), \\
    \pi &= \mr{\pi}_a \mr{\bar{m}}^a +O(\rho), \qquad&     \alpha &= \f{1}{2}(\mpi_a -  \mr{D}_a) \mr{\bar{m}}^a +O(\rho), \\
    \mu &= \f{1}{2}\mtn +O(\rho), \qquad& \beta &= \f{1}{2}(\mpi_a + \mr{D}_a) \mr{m}^a  + O(\rho), \\
    \lambda &= \mr{K}^{(n)}_{ab} \mr{\bar{m}}^a \mr{\bar{m}}^b +O(\rho), \qquad& \sigma &= -\mr{K}^{(\ell)}_{ab} \mr{m}^a \mr{m}^b + O(\rho),\\
    \epsilon &= \f{1}{2}\Bigl(\mr{\kappa}_{(\ell)}  +\f{1}{2}\mr{K}_{ab}^{(\ell)}(\mr{\bar{m}}^a\mr{\bar{m}}^b - \mr{m}^a\mr{m}^b) \Bigl)+ O(\rho) ,  \qquad& \nu&=\gamma=\tau=0.\\
\end{aligned}
\label{list_sc}
\end{equation}
To conclude the geometric characterization of $\scr{H}$, it was argued in \cite{Chandrasekaran:2021hxc} that, in order to define the analogue of the Brown-York tensor on a null hypersurface, it is necessary to introduce the Weingarten map, which describes how $\scr{H}$ bends in $\scr{M}$ and it is defined as follows
\begin{equation}
\cW_i^{\ j} := \Pi^\mu_{\ i}\nabla_\mu\ell^\nu \Pi_{\ \nu}^j.
\end{equation}
Then, the null Brown-York tensor, or the Carrollian stress-energy tensor, is defined as follows
\begin{equation}
    T_i^{\ j} =\cW_i^{\ j} -\var_i^{\ j}\cW_k^{\ k},
\end{equation}
and can be decomposed into three blocks, specifically
\begin{equation}
\begin{aligned}
\thell := -T^i_{\ j}\ \ell^j n_i,\qquad
\tau_a := T^i_{\ j}\ n_i\ q^j_{\ a}, \qquad
\tau^a_{\ b} :=T^i_{\ j}\ q^j_{\ b} q^a_{\ i}.
\end{aligned}
\end{equation}
In terms of the previously defined geometric quantities, they yield
\begin{equation}
\tau_a = \f{1}{8\pi G}\Bigl(\pi_a - \thell n_a\Bigl), \qquad \text{and} \qquad \tau^a_{\ b} = \f{1}{8\pi G}\Bigl(\sigma^{(\ell)a}_{b} -\mu\var_{\ b}^a\Bigl).
\end{equation}
Now that we have characterized the geometry on a null horizon, in the next subsection we discuss the symmetry properties of $\scr{H}$.

\subsection{Covariant functionals}
In \cite{DeSimone:2026nig} it was shown that it is possible to construct a set of boost-weighted functionals of the metric which transform covariantly under the action of the homogeneous subgroup $H_S := \Diff(\scr{S}) \ltimes \R_W$ of the near-horizon symmetry group, which fully consists of $(\Diff(\scr{S}) \ltimes \R_W) \ltimes \R_T$ \cite{Chandrasekaran:2018aop}. Specifically, $H_S$ consists of a semi-direct product between $\scr{S}$-diffeomorphisms and Weyl super-boosts, while the full near horizon group also includes super-translations. These transformations are parametrized by three nowhere vanishing functions defined on $\scr{S}$, denoted by $Y^a(z), \ W(z)$ and $T(z)$ respectively.\\
In particular, we say that a quantity $\cO$ transforms with a definite boost weight $w$ under \eqref{resc_ln} if it obeys the following transformation rule
\begin{equation}
    \cO_{(w)} \to e^{w\lambda_L}\cO_{(w)},
\end{equation}
and we say that $\cO$ transforms semi-covariantly under the action of the full near horizon symmetry group if it manifests the following behaviour
\begin{equation}
    \var \cO_{(w)} = (\tau\pa_v + \Lie_Y + w\dot{\tau})\cO_{(w)} - L^a_{\cO} \pa_a\tau,
\end{equation}
where $\tau=T+vW$ and $L^a_{\cO}$ is the linear anomaly of $\cO$. These boost-weighted semi-covariant functionals are constructed out of the geometric fields defined in the previous section (see \cite{DeSimone:2026nig} for details) and by contracting them with the complex dyad, one defines the following scalars
\begin{equation}
\begin{aligned}
\cN:=\cN_{ab}\ \mr{m}^a\mr{m}^b,& \qquad \cJ:=\cJ_{a}\ \mr{m}^a, \qquad \cA:=\cA_{ab}\ \mr{m}^a\mr{\bar{m}}^b= \cA_{\Re} + i\cA_{\Im},\\
&\cP:=\cP_{a}\ \mr{\bar{m}}^a, \qquad \cT:=\cT_{ab}\ \mr{\bar{m}}^a\mr{\bar{m}}^b,    
\end{aligned}
\label{obs}
\end{equation}
where $w(\cN)=2, \ w(\cJ)=1, \ w(\cA)=0, \ w(\cP)=-1$ and $w(\cT)=-2$. In particular, these quantities represent the leading order of the Weyl scalars, i.e.,
\begin{equation}
\begin{aligned}
{\Psi}_{4} &=- \cT +\rho \Psi_{4}^{(1)} + o(\rho), \\
{\Psi}_{3} &= - \cP + \rho\Bigl( \eth \cT - \mr{\mu}\cP\Bigl) + o(\rho),\\
{\Psi}_{2} &= -\cA + \rho\Bigl(\eth\cP -3\mr{\mu}\cA + \mr{\sigma}\cT \Bigl) + o(\rho), \\
{\Psi}_{1} &= -\cJ + \rho \Bigl( \eth\cA -2\mr{\mu}\cJ +2\mr{\sigma} \cP \Bigl) + o(\rho), \\
\Psi_{0} &= -\cN + \rho\Bigl( \eth \cJ -\mr{\mu}\cN + 3\mr{\sigma}\cA \Bigl) + o(\rho).
\end{aligned}
\end{equation}
As shown in \cite{DeSimone:2026nig}, the knowledge of the boost-weighted $w = -2$ functional and its symmetry property, allows us to derive the expression of the boost-weighted $w = -1$ functional; the knowledge of the $w = -1$ functional to derive the $w =0$ functional, and so on. In other words, we have the following pattern
\begin{equation}
\begin{aligned}
\var_{(\tau,Y)}\cT_{ab} &= (\tau\pa_v +\Lie_Y -2\dot{\tau})\cT_{ab} + 4\cP_{\langle a} \pa_{b \rangle}\tau,\\
\var_{(\tau, Y)}\cP_a &\heq (\tau\pa_v + \Lie_Y - \dot{\tau}) \cP_a + 3\cA_{ab}\pa^b\tau.\\
\var_\xi \cA_{\Re}&=(\tau\pa_v + \Lie_Y)\cA_{\Re}  + 2\cJ^a \pa_a \tau,\\
\var_\xi \cA_{\Im} &= (\tau\pa_v + \Lie_Y) {\cA}_{\Im}  + 2\varepsilon^{ab}{\cJ}_b \pa_a \tau,\\
\var_\xi \cJ_a &= (\tau\pa_v +\Lie_Y + \dot{\tau}) \cJ_a + \cN_a^{\ b}\pa_b\tau,\\
\var_\xi\cN_{ab} &= (\tau\pa_v + \Lie_Y +2\dot{\tau}) \cN_{ab},
\end{aligned}
\label{ite_bw}
\end{equation}
where $\varepsilon_{ab}=\epsilon_{ab}\sqrt{\mr{q}}$. A recursive formula in terms of the boost weight is given in \cite{DeSimone:2026nig}. However, in this work, we mainly work with the concept of spin weight $s$, defined in the next subsection and, therefore, the recursive formula coming from \eqref{ite_bw} will be given in terms of $s$.

\subsection{Dynamics and weighted operators}
Before delving into the discussion of the near-horizon dynamics, let us briefly introduce a  powerful tool to analyse null geometries through the spin formalism, introduced by Penrose, Newman, Geroch, Rindler and others \cite{Newman:1961qr, Geroch:1968zm, Geroch:1970uv, Geroch:1973am, Penrose:1985bww}. This approach is fundamental for revealing a recursive pattern in the dynamics of arbitrary null hypersurfaces. Let us define a spin frame $(o^A,\iota^A)$
\begin{equation}
    \ell^\mu = o^A\bar{o}^{A'}, \qquad n^\mu = \iota^A\bar{\iota}^{A'}, \qquad m^\mu = o^A\bar{\iota}^{A'}, \qquad \bar{m}^\mu = \iota^A\bar{o}^{A'}
\end{equation}
which is uniquely defined up to the sign, $(o^A,\iota^A) \to (-o^A,-\iota^A)$. The directional derivatives are defined as \footnote{
In particular, GHP formalism uses only six Greek letter instead of the twelve of the NP formalism. The relation between the two formalism is
\begin{equation}
\begin{aligned}
\nu = -\kappa', \qquad \lambda=-\sigma', \qquad \mu = -\varrho'\\
    \pi = -\tau', \qquad \alpha = -\beta', \qquad \gamma = -\varepsilon'
\end{aligned}
\end{equation}
}
\begin{equation}
\begin{aligned}
\mathbb{D}=o^A \bar{o}^{A'}\nabla_{AA'} = \ell^\mu\nabla_\mu, \qquad \mathbb{D}'=\mathbb{\Delta}=\iota^A \bar{\iota}^{A'}\nabla_{AA'} = n^\mu\nabla_\mu,\\
\vardelta=o^A \bar{\iota}^{A'}\nabla_{AA'} = m^\mu\nabla_\mu, \qquad \vardelta' = \bar{\vardelta}=\iota^A \bar{o}^{A'}\nabla_{AA'} = \bar{m}^\mu\nabla_\mu. 
\end{aligned}
\label{dir_der}
\end{equation}
By setting $o^A\iota_A=1$, the following rescaling leaves the null directions invariant
\begin{equation}
o^A\to \mathbb{\Lambda} o^A, \qquad \iota^A\to \mathbb{\Lambda}^{-1}\iota^A,
    \label{resc_oa}
\end{equation}
where $\mathbb{\Lambda}$ is a nowhere vanishing complex scalar field. Taking $\mathbb{\Lambda}=e^{\f{1}{2}(\lambda_L+i\vartheta)}$, where $\lambda_L$ and $\vartheta$ are arbitrary real functions, the rescaling \eqref{resc_oa} is nothing other than the Lorentz transformations of class III, since the tetrad transforms as follows
\begin{equation}
\ell\to e^{\lambda_L}\ell, \qquad n\to e^{-\lambda_L}n,\qquad m\to e^{i\vartheta}m, \qquad \bar{m}\to e^{-i\vartheta}\bar{m}. \label{resc_tetr}
\end{equation}
Therefore, a quantity $\eta$ is of type $\{p,q\}$ if it transforms as follows 
\begin{equation}
    \eta \to \mathbb{\Lambda}^{p}\bar{\mathbb{\Lambda}}^{q}\eta
\end{equation}
under \eqref{resc_oa}. Equivalently, we say that $\eta$ has spin weight \footnote{Actually, the convention usually adopted for the spin is $s=(p-q)/2$. Here, we take the opposite sign for a comparison with the asymptotic results.}  $s=(q-p)/2$ and boost weight $w=(p+q)/2$. \\
However, when the directional derivatives in \eqref{dir_der} are applied to weighted scalars, they do not in general produce a ${p,q}$ weighted quantity. Therefore, one has to modify the derivative operators by including the spin-coefficients that are not weighted, namely $\varepsilon, \gamma, \alpha$ and $\beta$. These $\{p,q\}$ weighted operators are defined as follows
\begin{equation}
\begin{aligned}
\pth \eta &= (\mathbb{D} - p\epsilon -q\bar{\epsilon} )\eta,\qquad  \pth' \eta = (\mathbb{\Delta} -p\gamma -q\bar{\gamma})\eta, \\
\eth \eta &= (\vardelta -p\beta -q\bar{\alpha})\eta, \qquad
\eth' \eta = (\bar{\vardelta} -p\alpha-q\bar{\beta} )\eta,\\
\end{aligned}
\label{sw_oper}
\end{equation}
and possess the following weights
\begin{equation}
\pth=\{1,1\}, \qquad \pth'=\{-1,-1\}, \qquad \eth=\{1,-1\}, \qquad \eth'=\{-1,1\}.
\end{equation}
Now, in terms of the spin weight $s$ we define the following spin-weighted charge aspects
\begin{equation}
\cQ_{-2}=\cN, \qquad \cQ_{-1}=\cJ, \qquad \cQ_{0}=\cA, \qquad \cQ_{1}=\cP, \qquad \cQ_{2}=\cT,
\end{equation}
and the equations \eqref{ite_bw} can be organized elegantly in the following recursive pattern
\begin{equation}
    \var_{(\tau, Y)}\cQ_{s} = (\tau\pa_v + \Lie_Y +w\dot{\tau}) \cQ_s +(s+2)\cQ_{s-1}\pa\tau.
\end{equation}
In addition to the boost and spin transformations, the null cone structure is also preserved under a Weyl rescaling of the metric $g_{\mu\nu}\to \mathbb{\Omega}^2 g_{\mu\nu}$. Under this transformation, the spinors transform as
\begin{equation}
o^A\to \mathbb{\Omega}^{\omega_0}o^A, \qquad \iota^A\to \mathbb{\Omega}^{\omega_1}\iota^A,
    \label{weyl_oa}
\end{equation}
and the tetrad as follows
\begin{equation}
\ell\to \mathbb{\Omega}^{2\omega_0}\ell, \qquad n\to \mathbb{\Omega}^{2\omega_1}n, \qquad m\to \mathbb{\Omega}^{\omega_0+\omega_1}m, \qquad \bar{m}\to \mathbb{\Omega}^{\omega_0+\omega_1}\bar{m},
    \label{weyl_tetr}
\end{equation}
where $\mathbb{\Omega}$ is a positive-definite scalar field and $\omega_{0,1}$ are the Weyl weights of the spinor fields. Then, we say that a quantity $\eta$ has Weyl weight $\omega$ if it transforms as follows
\begin{equation}
    \eta\to\mathbb{\Omega}^\omega \eta
\end{equation}
under \eqref{weyl_oa}. The weighted derivative operators defined in \eqref{sw_oper} can be further developed to accommodate Weyl weights, and become
\begin{equation}
\begin{aligned}
\pth_{\scr{C}}  &= \pth + [\omega + (p+q)\omega_1]\varrho,\qquad & \pth'_{\scr{C}} = \pth' -[\omega - (p+q)\omega_0]\mu, \\
\eth_{\scr{C}}  &= \eth + [\omega +p\omega_1 -q\omega_0]\tau, \qquad&
\eth'_{\scr{C}}  = \eth' - [\omega -p\omega_0 +q\omega_1]\pi,
\end{aligned}
\end{equation}
whose $\{p,q,\omega\}$-weights are
\begin{equation}
\begin{aligned}
    \pth_{\scr{C}}&=\{1,1,2\omega_0\}, \qquad& \pth'_{\scr{C}}=\{-1,-1, 2\omega_1\}, \\
    \eth_{\scr{C}}&=\{1,-1, \omega_0+\omega_1\}, \qquad& \eth'_{\scr{C}} = \{-1,1,\omega_0+\omega_1\}.
\end{aligned}
\end{equation}
Below is the list of the spin coefficients with well defined $\{p,q,\omega\}$ weights \footnote{The $\{p,q,\omega\}$ weights of the null frame are 
\begin{equation}
\ell = \{1,1,2\omega_0\}, \quad n= \{-1,-1,2\omega_1\}, \quad m= \{1,-1,\omega_0+\omega_1\}, \quad \bar{m}= \{-1,1,\omega_0+\omega_1\}.    
\end{equation}
}
\begin{equation}
\begin{aligned}
\sigma=\{3,-1,2\omega_0\}, &\qquad \kappa=\{3,1,3\omega_0-\omega_1\}, \\ 
\lambda=\{-3,1, 2\omega_1\}, &\qquad \nu=\{-3,-1,3\omega_1-\omega_0\}.
\end{aligned}
\end{equation}
Moreover, the Weyl scalars transform covariantly under Weyl rescaling, possessing the following $\{p, q,\omega\}$-weights
\begin{equation}
\begin{aligned}
\Psi_0=\{4,0,4\omega_0-1\},& \qquad \Psi_1=\{2,0,3\omega_0+\omega_1-1 \}, \qquad \Psi_2=\{0,0, 2\omega_0+2\omega_1-1\},\\
&\Psi_3=\{-2,0, \omega_0+3\omega_1-1\}, \qquad \Psi_4=\{-4,0,4\omega_1-1\}.
\end{aligned}
\end{equation}
Since the spin-weighted charges are the leading order of the Weyl scalars near horizon, their dynamics is encoded in the near-horizon Bianchi identities, that read
\begin{equation}
\begin{aligned}
\cE_\cJ&:=\pth_{\scr{C}}\cJ - \eth'_{\scr{C}}\cN,\qquad&
\cE_\cA&:=\pth_{\scr{C}}\cA - \eth'_{\scr{C}}\cJ +\lambda\cN,\\
\cE_\cP&:= \pth_{\scr{C}}\cP - \eth'_{\scr{C}}\cA + 2\lambda\cJ,\qquad &
\cE_\cT &:=\pth_{\scr{C}}\cT - \eth'_{\scr{C}}\cP + 3\lambda\cA,
\end{aligned}
\label{ev_eq}
\end{equation}
which are Weyl covariant and possess the following $\{p,q,\omega\}$-weights
\begin{equation}
\begin{aligned}
\cE_\cJ&=\{3,1,5\omega_0+\omega_1-1\}, \qquad &\cE_\cA&=\{1,1,4\omega_0+2\omega_1-1\},\\
\cE_\cP&=\{-1,1,3\omega_0 +3\omega_1-1\}, \qquad &\cE_\cT&=\{-3,1,2\omega_0 +4\omega_1-1\}.
\end{aligned}
\end{equation}
In terms of the spin-weight, the leading Bianchi identities can be written in the following recursive formula
\begin{equation}
\pth_\scr{C} \cQ_s = \eth'_{\scr{C}} \cQ_{s-1} - (s+1)\mr{\lambda}\cQ_{s-2}, \qquad \text{with}\ s=-2,-1,0,1,2,
\end{equation}
and possess the following behaviour under the action of the near-horizon symmetry group \cite{DeSimone:2026nig}
\begin{equation}
\var_{(\tau, Y)}\cE_{\cQ_s} = (\tau\pa_v + \Lie_Y +w\dot{\tau}) \cE_{\cQ_s} +(s+1) \cE_{\cQ_{s-1}} \pa\tau.
\end{equation}

\subsection{Phase space}

The canonical pre-symplectic potential on an arbitrary null hypersurface is equivalent to the null analogue of the Brown-York potential \cite{Chandrasekaran:2021hxc} and reads as follows
\begin{equation}
\Theta^c = \int_{\scr{H}}\Bigl( \f{1}{2} \tau^{ab}\var q_{ab} -\tau_a\var\ell^a\Bigl)\eps_{\scr{H}}.
\label{theta_metr}
\end{equation}
In this work, we are interested in the leading part of the pre-symplectic potential where the spin-1 sector, namely the term $\tau_a\ell^a$, does not appear due to the boundary conditions we used. Therefore, the (leading) canonical symplectic 2-form can be arranged as follows \cite{Ciambelli:2023mir}
\begin{equation}
\mr{\Omega}^c = \f{1}{8\pi G}\int_{\scr{H}}  \Bigl(\var (\f{1}{2}\mr{\Omega}\mr{\sigma}_{(\ell)}^{ab}) \curlywedge  \var \mr{q}_{ab}  - \var\mr{\mu}_{(\ell)} \curlywedge \var \mr{\Omega}\Bigl)\ \dext v \wedge \dext^2 z.
\end{equation}
At this point, we want to evaluate the variation in the spin-2 sector. Let us work with a complex dyad ($m_a, \bar{m}_a$), such that the corner metric can be written as
\begin{equation}
    q_{ab}= m_a\bar{m}_b + m_b\bar{m}_a.
\end{equation}
The variation of the frames can be written as follows
\begin{equation}
\var \mr{m}_a = \mr{\bar{m}}_a \Delta + \mr{m}_a\omega,\qquad
\var \mr{\bar{m}}_a = \mr{m}_a \bar{\Delta} + \mr{\bar{m}}_a \bar{\omega}\\
\label{48}
\end{equation}
and
\begin{equation}
\var \mr{m}^a = -\mr{\bar{m}}^a \Delta - \mr{m}^a\bar{\omega},\qquad \var \mr{\bar{m}}^a = -\mr{m}^a \bar{\Delta} - \mr{\bar{m}}^a\omega,
\label{49}
\end{equation}
where $\Delta$ and $\omega$ are complex 1-forms in the field space. Then, the variation of the corner metric takes the following form
\begin{equation}
\var \mr{q}_{ab} = 2\mr{m}_a\mr{m}_b\ \bar{\Delta} +2\mr{\bar{m}}_a \mr{\bar{m}}_b\ \Delta + 2\mr{q}_{ab} \Re{\omega}.
\end{equation}
In particular, since from the spin-coefficient equation we have $\pth_{\scr{C}}\mr{m}^a = \sigma\mr{\bar{m}}^a$, i.e.,
\begin{equation}
\pa_v \mr{m}^a = \sigma \mr{\bar{m}}^a + (\mr{\varrho}+ \mr{\varepsilon} - \mr{\bar{\varepsilon}})\mr{m}^a,
\end{equation}
the field-space contraction of the \eqref{49} with $I_{\hat{v}}$ yields
\begin{equation}
I_{\hat{v}}\Delta = -\mr{\sigma}, \qquad I_{\hat{v}}\mr{\omega} = \mr{\varrho} + \mr{\varepsilon} - \mr{\bar{\varepsilon}}.
\end{equation}
Let us now evaluate the variation of the longitudinal shear. In particular, from the Newman-Penrose formalism we can write the longitudinal shear as
\begin{equation}
\mr{\sigma}^{(\ell)}_{ab} = -\mr{\sigma} \mr{\bar{m}}_a \mr{\bar{m}}_b - \mr{\bar{\sigma}} \mr{m}_a \mr{m}_b.
\end{equation}
Hence, using the equations \eqref{48} and \eqref{49} from the frame variation, we obtain
\begin{equation}
\begin{aligned}
\var\mr{\sigma}^{(\ell)}_{ab}&= -(\var+2\bar{\omega}) \mr{\sigma}\ \mr{\bar{m}}_a \mr{\bar{m}}_b - (\var+2\omega)\mr{\bar{\sigma}}\ \mr{m}_a \mr{m}_b - \mr{q}_{ab}(\mr{\sigma} \bar{\Delta} + \mr{\bar{\sigma}}\Delta)  \\
\end{aligned}
\end{equation}
and
\begin{equation}
\begin{aligned}
\var\mr{\sigma}^{ab}_{(\ell)} &= - (\var -2\omega)\mr{\sigma}\ \mr{\bar{m}}^a\mr{\bar{m}}^b - (\var -2\bar{\omega}) \mr{\bar{\sigma}}\ \mr{m}^a \mr{m}^b  + \mr{q}^{ab}(\mr{\sigma} \bar{\Delta} +\mr{\bar{\sigma}}\Delta). \\
\end{aligned}
\end{equation}
Putting all the previous contributions together, the spin-2 sector reads
\begin{equation}
\begin{aligned}
\mr{\sigma}^{ab}_{(\ell)}\var \mr{\Omega} \curlywedge  \var \mr{q}_{ab} + \mr{\Omega}\var \mr{\sigma}^{ab}_{(\ell)} \curlywedge  \var \mr{q}_{ab}&= -2\var \mr{\Omega} \curlywedge \mr{\sigma} \bar{\Delta} - \mr{\Omega}\Bigl(   2(\var -2\omega)\mr{\sigma}\ \curlywedge  \bar{\Delta} \\
& - 4\mr{\sigma} \bar{\Delta}\curlywedge  \Re{\omega}\Bigl) +\ \text{c.c.},
\end{aligned} 
\end{equation}
and substituting the above result in the symplectic 2-form, we finally obtain
\begin{equation}
\begin{aligned}
\mr{\Omega}^c &= -\f{1}{16\pi G}\int_{\scr{H}}  \Bigl(\mr{\sigma} \var \mr{\Omega} \curlywedge \bar{\Delta} + \mr{\Omega}\Bigl(   (\var -2\omega)\mr{\sigma}\ \curlywedge  \bar{\Delta}  - 2\mr{\sigma} \bar{\Delta}\curlywedge  \Re{\omega}\Bigl) +\ \text{c.c.} \\
&\qquad+ 2\var\mr{\mu}_{(\ell)} \curlywedge \var \mr{\Omega}\Bigl)\ \dext v \wedge \dext^2 z.
\end{aligned}
\end{equation}
The Poisson bracket can be extracted by following the computation presented in \cite{Ciambelli:2023mir}. However, our goal is to obtain a symplectic 2-form that resembles the structure of the asymptotic Ashtekar-Streubel phase space. Moreover, as in \cite{Freidel:2025ous} we assume the following decomposition of the boundary metric
\begin{equation}
    q_{ab} = \Omega(\gamma_{ab} + \varkappa h_{ab})
\end{equation}
where $\Omega$ is the area element, $\varkappa = \sqrt{32\pi G}$ and $\gamma_{ab}$ is some fixed boundary metric with $\var\gamma_{ab}=\pa_v\gamma_{ab}=0$. In this case, we have $\sigma_{a}^{(\ell)b} = \f{1}{2}\varkappa\pa_v h_{a}^{\ b}$. This decomposition turns out to be convenient in later discussion.

\subsection{Dressing time}
The structure of the near-horizon Bianchi identities is much more complicated with respect to the asymptotically flat analogue. Indeed, the integration of the \eqref{ev_eq} is not as straightforward as in the asymptotic flat case and one cannot compute the higher spin bracket by directly using the same methodology performed in \cite{Freidel:2021dfs, Freidel:2021ytz, Geiller:2024bgf}. To handle this problem, in the next section we define a new dressing time reference frame. Before delving into our construction, let us recall for comparison some dressing time proposals presented in the literature \cite{Ciambelli:2023mir, Ciambelli:2024swv, Freidel:2025ous, Ruzziconi:2025fuy}.

In \cite{Ruzziconi:2025fuy} the authors defined a dressing time via the condition $\pth_{\scr{C}} V^{\RZ}=1$, where $V^\RZ$ has weights $\{-1, -1, -2\omega_0\}$. Assuming $\omega_0=0$ and $\omega_1=-1$, explicitly we have
\begin{equation}
\pa_vV^{\RZ} =1 -(\mr{\kappa}_{(\ell)} -\mthell)V^{\RZ},
\end{equation}
whose general solution is
\begin{equation}
V^{\RZ} =  e^{-\int (\mr{\kappa}_{(\ell)} -\mthell)\dext v} \Bigl( V_0 +\int \dext v'\ e^{\int (\mr{\kappa}_{(\ell)} -\mthell)\dext v}\Bigl).
\end{equation}
Therefore, one could infer that in the RZ-dressing time reference frame the degrees of freedom encoded into the spin-coefficients $\varepsilon$ and $\varrho$ are absorbed into the dynamical coordinate $V^\RZ$.\\
On the other hand, the authors in \cite{Ciambelli:2023mir, Ciambelli:2024swv, Freidel:2025ous} define the dressing time as the reference frame where the boost connection vanishes. In other words, they define a combination of boost+diffeomorphism $V: v\to V$, under which
\begin{equation}
\mu_{(\ell)}= \pa_v V(\tilde{\mu}_{(\ell)} \circ V) +\frac{\pa_v^2 V}{\pa_vV},
\end{equation}
so that the equation that defines the dressing time frame is
\begin{equation}
    (\pa_v -\mu_{(\ell)})\pa_vV^{\CFL}=0,
\label{CFL_eq}
\end{equation}
whose solution takes the following form
\begin{equation}
V^{\CFL}_{(ab)} = V^{\CFL}(a) + \pa_v V^{\CFL}(b) \int_a^v\dext v' \ e^{\int_b^{v'} \dext v''\ \mu_{(\ell)}} (v'').
\end{equation}
Using the Newman-Penrose formalism, the CFL dressing time can be derived by performing a class III transformation with $\vartheta=0$ (see appendix \ref{weyl+III}) and requiring $\mr{\varepsilon}' + \mr{\bar{\varepsilon}}' - \mr{\varrho}'=0$, from which
\begin{equation}
\lambda_L=-\int \dext v\ (\mr{\varepsilon}+\mr{\bar{\varepsilon}} -\mr{\varrho}).
\end{equation}
Since $\exp{-\lambda_L}=\pa_vV$, then we exactly obtain the \eqref{CFL_eq}. \\
The CFL dressing time defines a dynamical reference frame, according to the construction in \cite{Goeller:2022rsx}. Indeed, the CFL dressing time has been determined via a field-dependent diffeomorphism and reflects the relational construction employed in \cite{Carrozza:2021gju, Goeller:2022rsx}. On the other hand, at first sight the RZ dressing time does not seem to follow the construction of \cite{Goeller:2022rsx}. It will be interesting to see if the RZ dressing time can be derived via field-dependent diffeomorphisms and whether this establishes a relation with the CFL dressing time.\\

\section{Dynamical reference frames}\label{sec2}
In the previous section, we argued that the Bianchi identities on an arbitrary null hypersurface are much more complicated than the asymptotic null analogue. These complications stem from the presence of boundary degrees of freedom in the former case, which are absent in the latter case due to the imposition of the boundary conditions on $\scri^+$. This leads to the replacement of $\pa_v\to\pth_{\scr{C}}$ and $\eth\to\eth'_{\scr{C}}$.

If one wants to mimic the methodology used in \cite{Freidel:2021dfs, Freidel:2021ytz, Geiller:2024bgf}, the introduction of the inverse operator $\pth_{\scr{C}}^{-1}$ is needed. However, this is a non trivial task due to the expression of the operator $\pth_{\scr{C}}$. Moreover, the structure of the symplectic form on $\scr{H}$ differs drastically from the Ashtekar-Streubel symplectic structure at null infinity. In this section, we address these two problems. Firstly, we saw that the introduction of the CFL dressing time cancels out the spin-0 sector in the symplectic form. Therefore, one can use the boost transformations to define a dynamical reference frame where the (leading term) of the longitudinal expansion $\mr{\varrho}$ is set to zero, instead of the surface tension. However, this is not enough because we still have the spin coefficient $\mr{\varepsilon}$ that enters into the definition of $\pth'_{\scr{C}}$. Most importantly, we want to preserve the structure of the Bianchi identities by looking for transformations under which the Bianchi identities themselves transform covariantly. As presented in the previous section, the Bianchi identities transform covariantly under Weyl rescaling and we use these transformations to set $\mr{\varepsilon}=0$. Similarly to \cite{Ciambelli:2024swv}, this allows us to define a dressing/dynamical reference frame $\tilde{\scr{M}}$ where $\tilde{\pth}_{\scr{C}}= \pa_{\tilde{v}}$, where $\tilde{v}$ is our dressing time. We dubbed this dynamical reference frame the \emph{Weyl reference frame}. It is worth-noticing that the null Raychaudhuri equation is not Weyl-covariant and, therefore, the requirement that $\mr{\tilde{\varrho}}=0 = \mr{\tilde{\varepsilon}}$ does not imply that $\mr{\tilde{\sigma}}$ vanishes (see appendix \ref{weyl+III}). To lighten the notation, from now on we drop the symbol $\mathring{\ }$ for the quantities evaluated in the Weyl reference frame, namely we always consider such quantities evaluated at the horizon, i.e., $\mr{\tilde{\sigma}}=\tilde{\sigma}$.

\subsection{Embedding field as DRF}\label{sec_2.1}
The choice of the embedding field is equivalent to the choice of a dynamical reference frame. The embedding field construction was firstly introduced in \cite{Donnelly:2016auv} and further developed in \cite{Speranza:2017gxd, Freidel:2021dxw, Ciambelli:2021nmv} for the description of the edge modes emerging in bounded subregions of spacetime. In \cite{Carrozza:2021gju, Carrozza:2022xut, Goeller:2022rsx} it was pointed out that the embedding field is nothing else that the inverse map of the dressing map. Therefore, the embedding map defines a dynamical reference frame. Furthermore, the introduction of the embedding map, hence the edge modes, led to the definition of the extended phase space. Thus, let us recall some fundamental ingredients of the embedding field construction and the extended phase space. \\
Let $\scr{M}$ be our spacetime manifold and consider another 4-dimensional manifold $\tilde{\scr{M}}$, called the reference manifold. These two manifolds are related via the following smooth embedding map
\begin{equation}
\scr{X}: \tilde{\scr{M}} \to \scr{M},
\end{equation}
and let $\scr{R}=\scr{X}^{-1}$ be the inverse of the embedding map, such that $\scr{R}(\scr{X}(\tilde{x}))=\tilde{x}$.
Given an $n$-form $\alpha$ on $\scr{M}$, we can define the relative $n$-form $\tilde{\alpha}$ on $\tilde{\scr{M}}$ via the pull-back map,
\begin{equation}
\scr{X}^*: \wedge^n T^*\scr{M} \to \wedge^n T^*\tilde{\scr{M}}, \qquad \scr{X}^*(\alpha(x))= \tilde{\alpha}(\scr{X}^{-1}(x))
\label{69}
\end{equation}
Let us see how the embedding map behaves under field-variation. Denoting by $\psi$ the field content of the theory, under a field-space variation we have
\begin{equation}
\begin{aligned}
\var(\scr{X}^*\alpha[\psi](x))
&= \scr{X}^*\Bigl(\var\alpha[\psi] + \Lie_{\scr{\chi}}\alpha[\psi] \Bigl),
\end{aligned}
\end{equation}
where $\chi$ is the Maurer-Cartan variational form, defined as follows
\begin{equation}
\chi = \var \scr{X}\circ \scr{X}^{-1},\qquad \text{with}\qquad \var\chi + \f{1}{2}[\chi,\chi]=0,
\label{71}
\end{equation}
and whose components are
\begin{equation}
\chi^\mu (x)= -(J^{-1})^{\mu}_{\ \nu}\ \var \scr{R}^\nu(x), \qquad \text{with}\qquad J^{\mu}_{\ \nu} = (\pa\scr{R}(x)/\pa x)^{\mu}_{\ \nu}.
\label{chi^mu}
\end{equation}
Now, via the pull-back map in \eqref{69} one can define the dressed Lagrangian $\tilde{\bd{L}} = \scr{X}^*\bd{L}$, whose field variation yields
\begin{equation}
\var\tilde{\bd{L}} = \scr{X}^*(\var\bd{L} +\Lie_{\chi}\bd{L}) \heq \dext\scr{X}^*(\bd{\theta} +\iota_{\chi}\bd{L}),
\end{equation}
so that the pre-symplectic potential associated with the dressed Lagrangian $\tilde{\bd{L}}$ is
\begin{equation}
\bd{\tilde{\theta}}[\psi, \var\psi] = \scr{X}^*\bd{\theta}_\chi[\psi, \var\psi]  = \scr{X}^*\Bigl(\bd{\theta}[\psi, \var\psi] +\iota_\chi \bd{L}[\psi]\Bigl).
\end{equation}
At this point, we introduce the field-space connection -- firstly derived in \cite{Gomes:2016mwl}--
\begin{equation}
    \var_\chi := \var + \Lie_\chi,
\end{equation}
and $\var^2_\chi=0$ from the Maurer-Cartan equation in \eqref{71}. Then, following the extended phase space construction \cite{Freidel:2021dxw,Speranza:2017gxd,Ciambelli:2021nmv}, the extended pre-symplectic current is defined as follows
\begin{equation}
\bd{\omega}_\chi := \var_\chi\bd{\theta}_\chi = \bd{\omega}  + \dext\Bigl(\iota_\chi\bd{\theta} +\f{1}{2}\iota_\chi\iota_\chi\bd{L}\Bigl)
\label{76}
\end{equation}
where $\bd{\omega}=\var\bd{\theta}$, and integrating \eqref{76} on a Cauchy slice we obtain the extended pre-symplectic 2-form
\begin{equation}
\Omega_\chi = \int_{\scr{H}}\bd{\omega}_\chi= \Omega + \int_{\scr{S}} \Bigl(\iota_\chi\bd{\theta} +\f{1}{2}\iota_\chi\iota_\chi\bd{L}\Bigl).
\end{equation}
Therefore, in addition to the bulk pre-symplectic 2-form $\Omega$, a corner contribution emerges in the extended phase space construction, encoding information about the edge modes of the theory.
However, as we will argue in the next subsections, for the purposes of this work we only deal with the bulk part of the symplectic potential, leaving the analysis of the corner potential to a future work.

\subsection{Weyl reference frame}
The aim of this subsection is to define a dynamical frame in which the evolution Bianchi identities can be easily integrated. Therefore, we look for a field-dependent diffeomorphism that sets $\mr{\varepsilon}=0=\mr{\varrho}$ in a certain reference frame. 
As argued in section \ref{1}, the Bianchi identities are Weyl covariant. Therefore, we combine the class III transformations with a Weyl rescaling. Assuming $\omega_0=0$ and $\omega_1=-1$, the spin coefficients $\mr{\varrho}$ and $\mr{\varepsilon}$ transform as follows
\begin{equation}
\mr{\varepsilon}\to \tilde{\varepsilon}= e^{\lambda_L}\Bigl(\varepsilon +\pa_v \ln \mathbb{\Omega} +\f{1}{2}\mathbb{D} \lambda_L +\f{i}{2}\mathbb{D}\vartheta\Bigl), \qquad \mr{\varrho}\to \tilde{\varrho}=e^{\lambda_L}(\varrho -\pa_v\ln \mathbb{\Omega}).
\end{equation}
Then, by requiring $\tilde{\varrho}=0, \ \tilde{\varepsilon}+\tilde{\varepsilon}=0$ and $\tilde{\varepsilon}-\tilde{\varepsilon}=0$, we obtain
\begin{equation}
\mathbb{\Omega} = e^{\int \dext v\ \mr{\varrho}},\qquad \lambda_L= -\int \dext v\ (\mr{\varepsilon} +\mr{\bar{\varepsilon}} + 2\mr{\varrho}),\qquad
\vartheta= i \int \dext v\ (\mr{\varepsilon} -\mr{\bar{\varepsilon}}),
\end{equation}
from which we can extract the expression of the dressing time $\tilde{v}$, which is
\begin{equation}
\pa_v\tilde{v} = \exp{\int \dext v\ (\mr{\varepsilon} +\mr{\bar{\varepsilon}} + 2\mr{\varrho})},
\end{equation}
and the dressing time equation reads
\begin{equation}
(\pa_v - \mr{\kappa}_{(\ell)}+\mthell) \pa_v\tilde{v}=0.
\end{equation}
Hence, the dressing time $\tilde{v}$ defines a new (dynamical) local coordinate system in which the longitudinal expansion $\tilde{\varrho}$ and the spin coefficient $\tilde{\varepsilon}$ vanish on $\tilde{\scr{H}}$. It is straightforward to see that when the horizon is non-expanding, the transformations above act as the identity transformation. However, contrarily to the non-expanding case, in the new system we have a non-vanishing longitudinal shear -- this is because the null Raychaudhuri is not Weyl-covariant. In this case the field space Maurer-Cartan form is
\begin{equation}
    \chi^\mu (x)= -(J^{-1})^{\mu}_{\ \nu}\ \var \scr{R}^\nu(x),\label{chi^mu}
\end{equation}
where $J^{\mu}_{\ \nu} = (\pa\scr{R}(x)/\pa x)^{\mu}_{\ \nu}$ is the Jacobian transformation from the coordinates on $\scr{M}$ to the coordinates on $\tilde{\scr{M}}$.


\subsection{Poisson bracket}
By trading the leading order of the spin coefficients $\varrho$ and $\varepsilon$ for dynamical coordinates, we have defined a dynamical reference frame, which we call the \emph{Weyl reference frame}. In the Weyl reference frame we have that $\tilde{\pth}'_{\scr{C}} = \pa_{\tilde{v}}$. Therefore, the first task of our list is accomplished. \\
Now, we want to understand how the symplectic structure changes when moving to the Weyl reference frame. As argued in subsection \ref{sec_2.1}, the embedding map is indeed a dynamical reference frame and when extending the phase space to accommodate this embedding field, the symplectic 2-form acquires a corner term $\tilde{\Omega}^c_{\pa\tilde{\scr{H}}}$, i.e.,
\begin{equation}
\tilde{\Omega}^c = \tilde{\Omega}^c_{\tilde{\scr{H}}} +  \tilde{\Omega}^c_{\pa\tilde{\scr{H}}},
\end{equation}
which encodes the information about the edge modes of the theory. For our purposes, we only consider the bulk symplectic form $\tilde{\Omega}^c_{\scr{H}}$ and discard the corner contribution. However, the corner term represents the interplay between the two reference frames and is essential in the analysis of the edge modes. We left this analysis for future work.\\
So far, we worked in the most general setting -- that is the near-horizon Weyl BMS group \cite{Freidel:2021fxf}. Similarly to the analysis at null infinity where the authors in \cite{Freidel:2021ytz, Geiller:2024bgf} required that the field variation of the celestial sphere metric is zero, to obtain the higher spin algebra some conditions on the corner metric have to be imposed. Therefore, in order to obtain a suitable symplectic 2-form, we firstly require that
\begin{itemize}
    \item We assume the following behaviour of the frame fields under field variation
\begin{equation}
    \var \tilde{m}_a = \bar{\tilde{m}}_a \tilde{\Delta},\qquad \text{or}\qquad
\var \tilde{q}_{ab} = 2\bar{\tilde{\Delta}} \tilde{m}_a\tilde{m}_b + 2\tilde{\Delta} \bar{\tilde{m}}_a \bar{\tilde{m}}_b,
\end{equation}
in other words we impose $\tilde{\omega}=0$, which implies $\var\tilde{\Omega}=0$. Then, the field variation of the dressed shear is
\begin{equation}
\begin{aligned}
\var\tilde{\sigma}^{ab}&= - \var \tilde{\sigma}\ \bar{\tilde{m}}^a\bar{\tilde{m}}^b - \var \bar{\tilde{\sigma}}\ \tilde{m}^a \tilde{m}^b  + \tilde{q}^{ab}(\tilde{\sigma} \bar{\tilde{\Delta}} +\bar{\tilde{\sigma}} \tilde{\Delta}),
\end{aligned}
\end{equation}
and the spin-2 sector reads
\begin{equation}
\var\tilde{\sigma}^{ab}\curlywedge\var\tilde{q}_{ab}= -\var\tilde{\sigma} \curlywedge \tilde{\bar{\Delta}}  +c.c.
\end{equation}
\item Secondly, we assume that $\tilde{\Delta}$ can  be written as a total field-variation, i.e., 
\begin{equation}
    \tilde{\Delta} = -\f{1}{2}\var \tilde{h}.
\end{equation} 
\end{itemize}
\noindent
In particular, from the frame fields equations we also have $I_{\hat{v}} \tilde{\Delta}= -\f{1}{2}\pa_{\hat{v}} \tilde{h} = -\tilde{\sigma}$. Then, the canonical bulk symplectic 2-form finally reads
\begin{equation}
\tilde{\Omega}^c_{\scr{H}} = \f{1}{\varkappa^2} \int_{\scr{H}} \Bigl(\var\tilde{\sigma} \curlywedge\var \tilde{\bar{h}}  +c.c. \Bigl) \tilde{\Omega}\ \dext \tilde{v} \wedge \dext^2 \tilde{z}.
\label{AS_finite}
\end{equation}
The \eqref{AS_finite} can be thought of as the analogue of the Ashtekar-Streubel symplectic 2-form for generic null hypersurfaces. The Poisson bracket extracted from \eqref{AS_finite} takes the following form
\begin{equation}
\{\tilde{\sigma}(\tilde{v}, \tilde{z}), \tilde{\bar{h}}(\tilde{v}', \tilde{z}')\} = \varkappa^2\var(\tilde{v}-\tilde{v}')\var(\tilde{z},\tilde{z}'),
\label{fun_PB}
\end{equation}
where $\var(\tilde{z},\tilde{z}') = \tilde{\Omega}^{-1} \var^{(2)}(\tilde{z}-\tilde{z}')$.

\subsection{Non-radiative condition}
As argued in section \ref{1}, the Bianchi identities are Weyl-covariant and in the Weyl reference frame they assume the following form 
\begin{equation}
\begin{aligned}
\pa_{\tilde{v}} \tilde{\cQ}_{b,s} = \tilde{\eth}'_{\scr{C}} \tilde{\cQ}_{b,s+1} - (s+1)\tilde{\lambda}\tilde{\cQ}_{b,s+2},
\end{aligned}
\label{w_BI}
\end{equation}
where the subscript $b$ stands for `bare charges'. To understand why we introduced this subscript, let us firstly define the pseudo-differential operator
\begin{equation}
\pa^{-1}_{\tilde{v}}:= \int_{+\infty}^{\tilde{v}} \dext \tilde{v}_1 \int_{+\infty}^{\tilde{v}_1}\dext \tilde{v}_2\cdots \int_{+\infty}^{\tilde{v}_{n-1}}\dext \tilde{v}_n.
\end{equation}
Then, the integration of the Bianchi identities yields
\begin{equation}
\begin{aligned}
\tilde{\cJ}_b &= \pa_{\tilde{v}}^{-1}\tilde{\eth}'_{\scr{C}}\tilde{\cN},\\
\tilde{\cA}_b &= \pa_{\tilde{v}}^{-1}\tilde{\eth}'_{\scr{C}}\tilde{\cJ}_b - \pa_{\tilde{v}}^{-1}(\tilde{\lambda}\tilde{\cN}),\\
\tilde{\cP}_b &= \pa_{\tilde{v}}^{-1}\tilde{\eth}'_{\scr{C}}\tilde{\cA}_b - 2\pa_{\tilde{v}}^{-1}(\tilde{\lambda}\tilde{\cJ}_b),\\
\tilde{\cT}_b &= \pa_{\tilde{v}}^{-1}\tilde{\eth}'_{\scr{C}}\tilde{\cP}_b- 3\pa_{\tilde{v}}^{-1}(\tilde{\lambda} \tilde{\cA}_b),
\end{aligned}
\label{int_BI}
\end{equation}
where the following boundary conditions
\begin{equation}
    \tilde{\sigma}= O(|\tilde{v}|^{-1-s-\alpha}), \qquad \lim_{\tilde{v}\to+\infty}\tilde{\cQ}_{b,s}=0,\qquad \text{with} \ \alpha>0,
\label{fall_off_sigm}
\end{equation}
are assumed in order to integrate the \eqref{w_BI} for all spin $s$. However, these boundary conditions are not sufficient to fully determine the asymptotic behaviour of the system. Contrarily to the asymptotic flat case, further fall-off conditions need to be imposed on the following boundary data $\tilde{\lambda}, \tilde{\mu}$ and $\tilde{\pi}$. From \eqref{fall_off_sigm}, we have $\tilde{h}= O(|\tilde{v}|^{-s-\alpha})$ and we further impose
\begin{equation}
\tilde{\pi} = O(|\tilde{v}|^{-1-s-\beta}), \qquad \tilde{\mu} = \tilde{\mu}_{\infty}(\tilde{z})+ O(|\tilde{v}|^{-1-s-\alpha}), \qquad \tilde{\lambda} = O(|\tilde{v}|^{-s-\alpha}),
\label{fall_off_rest}
\end{equation}
with $\beta>\alpha$. Assuming that at $\tilde{v}\to+\infty$ the geometry stabilizes in a spherical configuration, $\tilde{\mu}_\infty$ does not depend on the variables $\tilde{z}$ and, therefore, is a constant. Moreover, from the fall off conditions \eqref{fall_off_rest}, we also have that the Weyl covariant derivative $\tilde{\eth}'_{\scr{C}\tilde{z}}$ becomes $\tilde{v}$-independent as $\tilde{v}\to\infty$, with decay rate $O(|\tilde{v}|^{1-s-\beta})$.
These conditions will be fundamental in the evaluation of the charge bracket.\\
The equations \eqref{int_BI} cannot be worked out in the same manner as the asymptotic case, due to the fact that the Weyl covariant derivative $\tilde{\eth}'_{\scr{C}}$ is time-dependent and therefore does not commute with $\pa^{-1}_{\tilde{v}}$. This non-commutativity introduces non-local terms in \eqref{int_BI} and, consequently, non-local terms in the Poisson brackets. This issue is handled in the next subsection, where the bare charges will be dressed by these non-local terms.\\
To conclude this subsection, it must be noticed that the NP equation that defines $\Psi_0$, i.e.,
\begin{equation}
    D\sigma -\var\kappa = (\varrho+\bar{\varrho})\sigma +(3\varepsilon -\bar{\varepsilon}) \sigma -(\tau-\bar{\pi} +\bar{\alpha}+3\beta)\kappa + \Psi_0,
\end{equation}
and which is not Weyl covariant, when evaluated at the leading order, since $\varrho= \bar{\varrho}$ and $\mr{\kappa}=0$, reads as follows
\begin{equation}
(\pa_v -3\mr{\varepsilon} +\mr{\bar{\varepsilon}} -2\mr{\varrho} ) \mr{\sigma} = \mr{\Psi}_0, \qquad\text{or} \qquad\pth_{\scr{C}} \mr{\sigma} = \mr{\Psi}_0,
\end{equation}
which is instead Weyl-covariant. Therefore, in the Weyl reference frame the non-radiative condition simply becomes
\begin{equation}
\tilde{\Psi}_0=0 \qquad \Rightarrow\qquad \pa_{\tilde{v}}\tilde{\sigma}(\tilde{v}, \tilde{z})=0,
\end{equation}
namely the dressed shear $\tilde{\sigma}$ does not depend on the dressing time variable $\tilde{v}$ when the non-radiation condition is imposed. This means that
\begin{equation}
    \tilde{h}(\tilde{v}, \tilde{z}) = \tilde{h}_0 (\tilde{z}) + \tilde{v}\tilde{\sigma}(\tilde{z}),
\end{equation}
where $\tilde{h}_0 (\tilde{z})$ is a smooth function on the corner. As we will see in the following subsection, the non-radiative condition that has to be considered to ensure that the charges are conserved in time is
\begin{equation}
    \tilde{\cN}=0=\tilde{\cJ}_b,
\end{equation}
as in the asymptotic null case \cite{Freidel:2021dfs, Freidel:2021ytz, Geiller:2024bgf}.

\subsection{Non-local charges}\label{Non-local charges}
As we argued above, contrarily to the asymptotic null case, the operator $\pa_{\tilde{v}}^{-1}$ and $\tilde{\eth}'_{\scr{C}}$ do not commute with each other, i.e. $[\pa_{\tilde{v}}^{-1},\tilde{\eth}'_{\scr{C}}]\neq0$, since the corner metric $\tilde{q}_{ab}$ depends on the dressing time $\tilde{v}$. Specifically, we have
\begin{equation}
[\pa_{\tilde{v}}^{-1},\tilde{\eth}'_{\scr{C}}] f :=\pa_{\tilde{v}}^{-1}\Delta^{(f)}(\tilde{v}',\tilde{v}, \tilde{z})
\label{long_mem}
\end{equation}
where $\Delta^{(f)}(\tilde{v}', \tilde{v}, \tilde{z})$ is defined in appendix \ref{app_B} and encodes the history of the corner metric. Therefore, it is a non-local term. Consequently, the computation of Poisson brackets gives rise to non-local terms. To avoid these complications - and to obtain a structure close to the asymptotic one - we absorb the \eqref{long_mem}-like non-local terms into the definition of the covariant functionals. After a bit of algebra, we obtain
\begin{equation}
\begin{aligned}
\tilde{\cJ} &= \tilde{\cJ}_b - \pa_{\tilde{v}}^{-1}\Delta^{(\tilde{\cN})},\\
\tilde{\cA} &= \tilde{\cA}_b - \pa_{\tilde{v}}^{-1}\Delta^{(\tilde{\cJ})} - \tilde{\eth}'_{\scr{C}} \pa_{\tilde{v}}^{-1}\Delta^{(\tilde{\sigma})}  - \pa_{\tilde{v}}^{-1}\tilde{\eth}'_{\scr{C}} \pa_{\tilde{v}}^{-1}\Delta^{(\tilde{\cN})} ,\\
\tilde{\cP} &= \tilde{\cP}_b -[\pa_{\tilde{v}}^{-1}, \tilde{\eth}_{\scr{C}}^{'3}]\pa_{\tilde{v}}^{-1}\tilde{\sigma}  - \pa_{\tilde{v}}^{-1}\tilde{\eth}'_{\scr{C}} \pa_{\tilde{v}}^{-1}\Delta^{(\tilde{\cJ})} - \pa_{\tilde{v}}^{-1}\tilde{\eth}^{'2}_{\scr{C}} \pa_{\tilde{v}}^{-1}\Delta^{(\tilde{\sigma})}\\
&\qquad  - \pa_{\tilde{v}}^{-1}\tilde{\eth}'_{\scr{C}}\pa_{\tilde{v}}^{-1}\tilde{\eth}'_{\scr{C}} \pa_{\tilde{v}}^{-1}\Delta^{(\tilde{\cN})} +2\pa_{\tilde{v}}^{-1}(\tilde{\lambda}\pa_{\tilde{v}}^{-1}\Delta^{(\tilde{\cN})}) +[\pa_{\tilde{v}}^{-1},\tilde{\eth}'_{\scr{C}}] \pa_{\tilde{v}}^{-1}(\tilde{\lambda} \pa_{\tilde{v}}\tilde{\sigma}),\\
\tilde{\cT} &= \tilde{\cT}_b -[\pa_{\tilde{v}}^{-1}, \tilde{\eth}'^{4}_{\scr{C}}] \pa_{\tilde{v}}^{-2}\tilde{\sigma} -\pa_{\tilde{v}}^{-1}\tilde{\eth}'_{\scr{C}}[\pa_{\tilde{v}}^{-1}, \tilde{\eth}_{\scr{C}}^{'3}]\pa_{\tilde{v}}^{-1}\tilde{\sigma}  - \pa_{\tilde{v}}^{-1}\tilde{\eth}'_{\scr{C}}\pa_{\tilde{v}}^{-1}\tilde{\eth}'_{\scr{C}} \pa_{\tilde{v}}^{-1}\Delta^{(\tilde{\cJ})} \\
&\qquad  - \pa_{\tilde{v}}^{-1}\tilde{\eth}'_{\scr{C}}\pa_{\tilde{v}}^{-1}\tilde{\eth}^{'2}_{\scr{C}} \pa_{\tilde{v}}^{-1}\Delta^{(\tilde{\sigma})} - \pa_{\tilde{v}}^{-1}\tilde{\eth}'_{\scr{C}}\pa_{\tilde{v}}^{-1}\tilde{\eth}'_{\scr{C}}\pa_{\tilde{v}}^{-1}\tilde{\eth}'_{\scr{C}} \pa_{\tilde{v}}^{-1}\Delta^{(\tilde{\cN})}  \\
&\qquad +2\pa_{\tilde{v}}^{-1}\tilde{\eth}'_{\scr{C}}\pa_{\tilde{v}}^{-1}(\tilde{\lambda}\pa_{\tilde{v}}^{-1}\Delta^{(\tilde{\cN})}) +\pa_{\tilde{v}}^{-1}\tilde{\eth}'_{\scr{C}}[\pa_{\tilde{v}}^{-1},\tilde{\eth}'_{\scr{C}}] \pa_{\tilde{v}}^{-1}(\tilde{\lambda} \pa_{\tilde{v}}\tilde{\sigma})
\\
&\qquad + 3\pa_{\tilde{v}}^{-1}(\tilde{\lambda}\pa_{\tilde{v}}^{-1}\Delta^{(\tilde{\cJ})}) + 3\pa_{\tilde{v}}^{-1}(\tilde{\lambda}\tilde{\eth}'_{\scr{C}} \pa_{\tilde{v}}^{-1}\Delta^{(\tilde{\sigma})}) + 3\pa_{\tilde{v}}^{-1}(\tilde{\lambda} \pa_{\tilde{v}}^{-1} \tilde{\eth}'_{\scr{C}} \pa_{\tilde{v}}^{-1}\Delta^{(\tilde{\cN})})\\
&\qquad+ [\pa_{\tilde{v}}^{-1}, \tilde{\eth}'_{\scr{C}}] \pa_{\tilde{v}}^{-1} (\tilde{\lambda}\tilde{\eth}'_{\scr{C}}\tilde{\sigma}),
\end{aligned}
\label{NL_charges}
\end{equation}
where for $s\geq0$ we also have memory terms where the transversal shear $\lambda$ enters. At first sight, the expressions for the non-local charges in \eqref{NL_charges} are quite messy. However, one can distinguish the following recursive pattern
\begin{equation}
\tilde{Q}_{s} = \tilde{Q}_{b, s} - \tilde{\cM}_s, 
\end{equation}
where we defined the memories as 
\begin{equation}
\begin{aligned}
\tilde{\cM}_s &= [\pa_{\tilde{v}}^{-1}, \tilde{\eth}'^{2+s}_{\scr{C}}] \pa_{\tilde{v}}^{-s}\tilde{\sigma} + \pa_{\tilde{v}}^{-1}\tilde{\eth}'_{\scr{C}}\tilde{\cM}_{s-1} - (s+1)\pa_{\tilde{v}}^{-1}(\tilde{\lambda}\tilde{\cM}_{s-2})\\
&\qquad - [\pa_{\tilde{v}}^{-1}, \tilde{\eth}'_{\scr{C}}] \pa_{\tilde{v}}^{-1} (\tilde{\lambda}\tilde{\cQ}_{s-3})
\end{aligned}
\end{equation}
with $\tilde{\cM}_{-1} = [\pa_{\tilde{v}}^{-1},\tilde{\eth}'_{\scr{C}}]\tilde{\cN}$ and $\tilde{\cM}_{-2}=0$. Furthermore, we also have
\begin{equation}
[\pa_{\tilde{v}}^{-1}, \tilde{\eth}'^n_{\scr{C}}] \tilde{\sigma} = \sum_{k=0}^{n-1} \tilde{\eth}'^k_{\scr{C}} \pa_{\tilde{v}}^{-1} \Delta^{(\tilde{\eth}'^{n-1-k}_{\scr{C}}\tilde{\sigma})}(\tilde{v}',\tilde{v}, \tilde{z}).
\end{equation}
Notably, this construction has been employed for mathematical convenience, but it would be highly interesting to understand the physical implications behind this procedure. Thus, trading the bare charges for the non-local charges, the integration of the Bianchi identities in \eqref{int_BI} takes the following familiar form 
\begin{equation}
\begin{aligned}
\tilde{\cN} &= \pa_{\tilde{v}}\tilde{\sigma},\\
\tilde{\cJ} &= \tilde{\eth}'_{\scr{C}} \tilde{\sigma},\\
\tilde{\cA} &= \tilde{\eth}_{\scr{C}}^{'2}\pa_{\tilde{v}}^{-1}\tilde{\sigma} -\pa_{\tilde{v}}^{-1}(\tilde{\lambda} \pa_{\tilde{v}}\tilde{\sigma}), \\
\tilde{\cP} &= \tilde{\eth}_{\scr{C}}^{'3}\pa_{\tilde{v}}^{-2}\tilde{\sigma} -\tilde{\eth}'_{\scr{C}}\pa_{\tilde{v}}^{-2}(\tilde{\lambda} \pa_{v}\tilde{\sigma})-2\pa_{\tilde{v}}^{-1}(\tilde{\lambda}\tilde{\eth}_{\scr{C}}' \tilde{\sigma}),\\
\tilde{\cT} &= \tilde{\eth}_{\scr{C}}^{'4}\pa_{\tilde{v}}^{-3}\tilde{\sigma} -\tilde{\eth}_{\scr{C}}^{'2}\pa_{\tilde{v}}^{-3}(\tilde{\lambda} \pa_{\tilde{v}} \tilde{\sigma}) - 2\tilde{\eth}'_{\scr{C}} \pa_{\tilde{v}}^{-2}  (\tilde{\lambda} \tilde{\eth}_{\scr{C}}' \tilde{\sigma}) -3\pa_{\tilde{v}}^{-1}(\tilde{\lambda} \pa_{\tilde{v}}^{-1}\tilde{\eth}_{\scr{C}}^{'2}\tilde{\sigma}) \\
&\qquad +3\pa_{\tilde{v}}^{-1}(\tilde{\lambda}\pa_{\tilde{v}}^{-1}(\tilde{\lambda} \pa_{\tilde{v}}\tilde{\sigma})).
\end{aligned}
\label{105}
\end{equation}
This is exactly the same structure obtained for asymptotically flat spacetimes \cite{Freidel:2021ytz, Geiller:2024bgf}. Then, we can expand each charge according to the number of oscillator fields it contains, i.e.,
\begin{equation}
\tilde{\cQ}_s =\sum_{k=1}^{\max[2,1+s]} \tilde{\cQ}_s^k.
\end{equation}
The linear contribution is
\begin{equation}
\tilde{\cQ}^1_{s\geq -2} = (\tilde{\eth}'_{\scr{C}} \pa_{\tilde{v}}^{-1})^{s+2}\pa_{\tilde{v}}\tilde{\sigma}
\end{equation}
and -- taking into account the order of the operators -- the quadratic contribution is
\begin{equation}
\tilde{\cQ}^2_{s\geq 0} = -\sum_{\ell=0}^{s} (\ell+1) (\tilde{\eth}'_{\scr{C}}\pa_{\tilde{v}}^{-1})^{s-\ell}\pa_{\tilde{v}}^{-1} (\tilde{\lambda} \pa_{\tilde{v}}(\pa_{\tilde{v}}^{-1}\tilde{\eth}'_{\scr{C}})^\ell \tilde{\sigma}).
\label{quad_ch}
\end{equation}
Now that we have the general expression for the linear and the quadratic charges, we can proceed by applying the same methodology discussed in \cite{Freidel:2021dfs, Freidel:2021ytz, Geiller:2024bgf}. Although the computations are quite similar, one has to be careful and to bear in mind that the dressing time derivative and the dressing covariant derivative on the corner do not commute with each other. In particular, from the fundamental Poisson bracket in \eqref{fun_PB} and \eqref{105}, follow straightforwardly the brackets between the $s=-2,-1$ non-local charges, that read
\begin{equation}
\begin{aligned}
\{\tilde{Q}_{-2}(\tilde{v}, \tilde{z}), \tilde{\bar{h}}(\tilde{v}', \tilde{z}')\} &= \{\pa_{\tilde{v}} \tilde{\sigma}(\tilde{v}, \tilde{z}), \tilde{\bar{h}}(\tilde{v}', \tilde{z}')\} = \varkappa^2 \pa_{\tilde{v}}\var(\tilde{v}-\tilde{v}')\var(\tilde{z},\tilde{z}'),\\
\{\tilde{Q}_{-1}(\tilde{v}, \tilde{z}), \tilde{\bar{h}}(\tilde{v}', \tilde{z}')\} &=\{\tilde{\eth}'_{\scr{C} \tilde{z}}\tilde{\sigma}(\tilde{v}, \tilde{z}), \tilde{\bar{h}}(\tilde{v}', \tilde{z}')\} = \varkappa^2 \var(\tilde{v}-\tilde{v}')\tilde{\eth}'_{\scr{C} \tilde{z}}\var(\tilde{z},\tilde{z}').
\end{aligned}
\label{2_1_action}
\end{equation}

\section{Higher spin bracket}\label{sec3}
In this section, we derive the action of the non-local  charges $\tilde{\cQ}_s$ on the gravitational radiative data $h_{ab}$, for $s\geq0$. To compute the action of these charges, we massively use the properties of the delta function, i.e.,
\begin{equation}
f(\tilde{v})\var(\tilde{v}-\tilde{v'}) = f(\tilde{v}') \var(\tilde{v}-\tilde{v'}), \qquad \pa_{\tilde{v}}^n \var(\tilde{v}-\tilde{v'}) = (-1)^n \pa^n_{\tilde{v}'}\var(\tilde{v}-\tilde{v}')
\end{equation}
-- where the first identity was used in \eqref{2_1_action} -- and pseudo-differential identities
\begin{equation}
\begin{aligned}
\pa^{-n}_{\tilde{v}} \Bigl(f({\tilde{v}}) \theta({\tilde{v}'-\tilde{v}})\Bigl) &= -(F({\tilde{v}'}) -F({\tilde{v}})) \theta({\tilde{v}'-\tilde{v}}),\\
\pa^{-n}_{\tilde{v}} \Bigl(f({\tilde{v}}) \delta({\tilde{v}-\tilde{v}'})\Bigl) &= -\f{({\tilde{v}-\tilde{v}'})^{n-1}}{(n-1)!} f(\tilde{v}')\theta({\tilde{v}'-\tilde{v}}),\\
\end{aligned}
\end{equation}
following exactly the same methodology used in \cite{Freidel:2021dfs, Freidel:2021ytz, Geiller:2024bgf}. Nonetheless, one has to be careful when evaluating the bracket because the Weyl covariant derivative $\tilde{\eth}'_{\scr{C}}$ depends on the dressing time $\tilde{v}$, contrarily to the asymptotic null case. Another identity that will come in handy during the bracket computation, is the Leibniz identity for pseudo-differential operator, that reads
\begin{equation}
\pa^{-1}_{\tilde{v}}(fg) = \sum_{n=0}^\infty (-1)^n (\pa^n_{\tilde{v}} f) (\pa_{\tilde{v}}^{-n-1}g),
\label{Leinb_1}
\end{equation}
from which follows another useful idenitity
\begin{equation}
\pa^{-1}_{\tilde{v}}\Bigl( \f{(-{\tilde{v}})^s}{s!} f({\tilde{v}})\Bigl) = \sum_{n=0}^{s} \f{(-{\tilde{v}})^n}{n!} \pa_{\tilde{v}}^{s-n-1} f(\tilde{v}).
\label{Leinb_2}
\end{equation}
Moreover, using the general Leibniz formula
\begin{equation}
\pa^{\alpha}_{\tilde{v}}(fg) = \sum_{n=0}^\infty \binom{\alpha}{n}(\pa^n_{\tilde{v}} f) (\pa_{\tilde{v}}^{\alpha-n}g), \qquad \text{with} \qquad \binom{\alpha}{n} = \f{(\alpha)_n}{n!}
\label{Leinb_gen}
\end{equation}
where $(\alpha)_n=\alpha(\alpha-1)\cdots(\alpha-n+1)$ is the falling factorial, the generalization of \eqref{Leinb_2} reads as follows
\begin{equation}
\pa^{\alpha}_{\tilde{v}}\Bigl( \f{{\tilde{v}}^k}{k!} f({\tilde{v}})\Bigl) = \sum_{n=0}^{k} \f{{\tilde{v}}^{k-n}}{(k-n)!} \frac{(\alpha)_n}{n!} \pa_{\tilde{v}}^{\alpha-n} f(\tilde{v}).
\label{Leinb_2gen}
\end{equation}
Then, introducing the operator $\Delta -1= \tilde{v}\pa_{\tilde{v}}$ and the function $g(\tilde{v}) =\pa_{\tilde{v}}^{\alpha-k} f(\tilde{v})$, the \eqref{Leinb_2gen} becomes
\begin{equation}
\pa^{\alpha}_{\tilde{v}}\Bigl( \f{{\tilde{v}}^k}{k!} f({\tilde{v}})\Bigl) = \f{1}{k!}\sum_{n=0}^{k} \binom{k}{n}(\alpha)_n\tilde{v}^{k-n}
\pa_{\tilde{v}}^{k-n} g(\tilde{v})= \f{1}{k!}\sum_{n=0}^{k} \binom{k}{n}(\alpha)_n (\Delta-1)_{k-n} g(\tilde{v}).
\end{equation}
Finally, using the Vandermonde identity \footnote{The Vandermonde identity is
\begin{equation}
    (x+y)_k = \sum_{n=0}^k \binom{k}{n}(x)_n (y)_{k-n}.
\end{equation}}
we obtain
\begin{equation}
\pa^{\alpha}_{\tilde{v}}\Bigl( \f{{\tilde{v}}^k}{k!} f({\tilde{v}})\Bigl) = \f{1}{k!}(\Delta+\alpha-1)_{k} \pa_{\tilde{v}}^{\alpha-k} f(\tilde{v}).
\label{Leinb_3gen}
\end{equation}
The computations performed in this section follow straightforwardly the same methodology used in \cite{Freidel:2021dfs, Freidel:2021ytz, Geiller:2024bgf} and are exactly the same when evaluating the Poisson bracket at linear level in the next section,
\begin{equation}
\{\tilde{q}_s(\tilde{z}), \tilde{q}_{s'}(\tilde{z}')\}^1 = \{\tilde{q}^1_s(\tilde{z}), \tilde{q}^2_{s'}(\tilde{z}')\} + \{\tilde{q}^2_s(\tilde{z}), \tilde{q}^1_{s'}(\tilde{z}')\},
\end{equation}
since here there is not $\tilde{v}$-dependence and therefore we refer the reader to the appendix E of \cite{Freidel:2021ytz}.

\subsection{$\tilde{\cQ}_0$ charge action}
In order to compute the $\tilde{Q}_0$-action, we use the fundamental Poisson brackets in \eqref{fun_PB} and \eqref{2_1_action} and the pseudo-differential identities listed in section \ref{sec3}. The non-local charge $\tilde{\cQ}_0$ can be decomposed into two contributions, which explicitly read
\begin{equation}
\tilde{\cQ}_0 (\tilde{v}, \tilde{z}) =\tilde{\cQ}_0^1 (\tilde{v}, \tilde{z}) +\tilde{\cQ}_0^2 (\tilde{v}, \tilde{z}) = \tilde{\eth}^{'2}_{\scr{C}}\pa^{-1}_{\tilde{v}}\tilde{\sigma}(\tilde{v}, \tilde{z}) -\pa^{-1}_{\tilde{v}}(\tilde{\lambda}\pa_{\tilde{v}}\tilde{\sigma})(\tilde{v}, \tilde{z}).
\end{equation}
The first term is linear in $\tilde{\sigma}$ --  similarly to the null case we can refer to it as the soft term -- and its action on $\bar{\tilde{h}}$ gives the following contribution
\begin{equation}
\begin{aligned}
\{\tilde{\cQ}^1_0(\tilde{v}, \tilde{z}), \bar{\tilde{h}}(\tilde{v}', \tilde{z}')\} &= \{\tilde{\eth}^{'2}_{\scr{C}\tilde{z}}\pa^{-1}_{\tilde{v}}\tilde{\sigma}(\tilde{v}, \tilde{z}), \bar{\tilde{h}}(\tilde{v}', \tilde{z}')\} \\
&= -\varkappa^2\theta(\tilde{v}'-\tilde{v})\tilde{\eth}^{'2}_{\scr{C}\tilde{z}}\var(\tilde{z},\tilde{z}').
\end{aligned}
\end{equation}
The second term is the analogue of the hard term in the asymptotic analysis and its action on $\bar{\tilde{h}}$ yields\footnote{The Poisson bracket $\{\bar{\tilde{h}}, \tilde{\lambda}\}=0$, while the bracket $\{\tilde{h},\tilde{\lambda}\}\neq0$.}
\begin{equation}
\begin{aligned}
\{\tilde{\cQ}_0^2(\tilde{v}, \tilde{z}), \bar{\tilde{h}}(\tilde{v}', \tilde{z}')\} &= - \{\pa^{-1}_{\tilde{v}}(\tilde{\lambda}\pa_{\tilde{v}}\tilde{\sigma})(\tilde{v}, \tilde{z}), \bar{\tilde{h}}(\tilde{v}', \tilde{z}')\}\\
&=-\varkappa^2\pa_{\tilde{v}'}\Bigl(\tilde{\lambda}(\tilde{v}',\tilde{z}) \theta(\tilde{v}'-\tilde{v}) \var(\tilde{z},\tilde{z}')\Bigl).
\end{aligned}
\end{equation}
Putting the previous contributions together, the total action of the non-local charge $\tilde{\cQ}_0$ on $\bar{\tilde{h}}$ reads as follows
\begin{equation}
\begin{aligned}
\{\tilde{\cQ}_0(\tilde{v}, \tilde{z}), \bar{\tilde{h}}(\tilde{v}', \tilde{z}')\}  &= -\varkappa^2\Bigl[\theta(\tilde{v}'-\tilde{v})\tilde{\eth}^{'2}_{\scr{C}\tilde{z}}\var(\tilde{z},\tilde{z}')\\
&\qquad +\pa_{\tilde{v}'}\Bigl(\tilde{\lambda}(\tilde{v}',\tilde{z}) \theta(\tilde{v}'-\tilde{v}) \var(\tilde{z},\tilde{z}')\Bigl)\Bigl],
\end{aligned}
\end{equation}
and for $\tilde{v}'>\tilde{v}$ we find
\begin{equation}
\begin{aligned}
\{\tilde{\cQ}_0(\tilde{v}, \tilde{z}), \bar{\tilde{h}}(\tilde{v}', \tilde{z}')\}  &= -\varkappa^2\Bigl[\tilde{\eth}^{'2}_{\scr{C}\tilde{z}}\var(\tilde{z},\tilde{z}')+\pa_{\tilde{v}'}\Bigl(\tilde{\lambda}(\tilde{v}',\tilde{z}) \var(\tilde{z},\tilde{z}')\Bigl)\Bigl].
\end{aligned}
\label{117}
\end{equation}
Now, we want to define the near-horizon generator of the spin-0 symmetry. Let us assume that there exists a smooth parameter $\tilde{\tau}_0(\tilde{z})$ defined on $\tilde{\scr{S}}$ such that 
\begin{equation}
\tilde{Q}_0 = \int_{\tilde{\scr{S}}} \tilde{\tau}_0(z)\tilde{\cQ}_0(\tilde{v},\tilde{z})\ \eps_{\tilde{\scr{S}}},
\end{equation}
and evaluating the symmetry transformation induced by $\tilde{Q}_0$ on $\bar{\tilde{h}}$ as $\tilde{v}\to+\infty$, we obtain
\begin{equation}
\begin{aligned}
\var_{\tilde{\tau}_0}\bar{\tilde{h}}(\tilde{v}\to+\infty, \tilde{z}) &= -\tilde{\eth}^{'2}_{\scr{C}\tilde{z}}\tilde{\tau}_0 -\tilde{\tau}_0 \pa_{\tilde{v}}\tilde{\lambda}(\tilde{v},\tilde{z})\Bigl\vert_{\tilde{v}\to+\infty}\\
&\heq -\tilde{\eth}^{'2}_{\scr{C}\tilde{z}}\tilde{\tau}_0 -\tilde{\tau}_0 \tilde{\mu}_{\infty}\bar{\tilde{\sigma}}(+\infty,\tilde{z}) \\ 
&= -\tilde{\eth}^{'2}_{\scr{C}\tilde{z}}\tilde{\tau}_0 +\f{\tilde{\mu}_{\infty}}{2}\tilde{\tau}_0 \pa_{\tilde{v}} \bar{\tilde{h}}\Bigl\vert_{\tilde{v}\to+\infty},
\end{aligned}
\end{equation}
where the fall-off condition \eqref{fall_off_sigm}-\eqref{fall_off_rest} have been used. Hence, the linear term of the $\tilde{Q}_0$-action yields a 'soft' contribution as expected, responsible for the near-horizon memory effect. Indeed, letting the geometry evolve as $\tilde{v}$ goes to $+\infty$, the analogue transformation found at null infinity emerges at finite distance \footnote{The asymptotic transversal shear $C$ is replaced by $\bar{\tilde{h}}$.}. Further and detailed investigations about the study of the memory effect in the Weyl reference frame are carried out in a forthcoming work.

\subsection{$\tilde{\cQ}_1$ charge action}
Let us now compute the action of the non-local spin-1 charge $\tilde{\cQ}_{1}$ on $\bar{\tilde{h}}$. Again, we can split the charge expression into three contributions
\begin{equation}
\tilde{\cQ}_{1}(\tilde{v}, \tilde{z}) = \tilde{\eth}^{'3}_{\scr{C}}\pa_{\tilde{v}}^{-2}\tilde{\sigma}(\tilde{v}, \tilde{z}) -\tilde{\eth}'_{\scr{C}}\pa_{\tilde{v}}^{-2}(\tilde{\lambda} \pa_{\tilde{v}}\tilde{\sigma})(\tilde{v}, \tilde{z}) -2\pa_{\tilde{v}}^{-1}(\tilde{\lambda} \tilde{\eth}'_{\scr{C}}\tilde{\sigma})(\tilde{v}, \tilde{z}).
\end{equation}
The first term, which represents the linear contribution, yields
\begin{equation}
\begin{aligned}
\{\tilde{\eth}^{'3}_{\scr{C}}\pa_{\tilde{v}}^{-2}\tilde{\sigma}(\tilde{v}, \tilde{z}), \bar{\tilde{h}}(\tilde{v}', \tilde{z}')\} &= - \varkappa^2 (\tilde{v} - \tilde{v}') \theta(\tilde{v}' - \tilde{v}) \tilde{\eth}^{'3}_{\scr{C}\tilde{z}}\var(\tilde{z},\tilde{z}').
\end{aligned}
\end{equation}
The Poisson bracket with the second and third terms represents the quadratic action, and gives the following contributions
\begin{equation}
\begin{aligned}
\{\tilde{\eth}'_{\scr{C}}\pa_{\tilde{v}}^{-2}(\tilde{\lambda} \pa_{\tilde{v}}\tilde{\sigma})(\tilde{v}, \tilde{z}), \bar{\tilde{h}}(\tilde{v}', \tilde{z}')\} &= \varkappa^2 \tilde{\eth}'_{\scr{C} \tilde{z}} \pa^{-2}_{\tilde{v}}[\tilde{\lambda}(\tilde{v}, \tilde{z}) \pa_{\tilde{v}} \var(\tilde{v} - \tilde{v}')\var(\tilde{z},\tilde{z}')]\\
&=\varkappa^2 \tilde{\eth}'_{\scr{C}\tilde{z}} \pa_{\tilde{v}'} \Bigl[(\tilde{v} - \tilde{v}')\tilde{\lambda}(\tilde{v}', \tilde{z}) \theta(\tilde{v}' - \tilde{v})\var(\tilde{z},\tilde{z}')\Bigl],
\end{aligned}
\end{equation}
and
\begin{equation}
\begin{aligned}
\{\pa^{-1}_{\tilde{v}}(\tilde{\lambda}\tilde{\eth}'_{\scr{C}\tilde{z}}\tilde{\sigma}) (\tilde{v}, \tilde{z}), \bar{\tilde{h}}(\tilde{v}', \tilde{z}')\} &= -\varkappa^2 \tilde{\lambda}(\tilde{v}', \tilde{z}) \theta(\tilde{v}' - \tilde{v}) \tilde{\eth}'_{\scr{C}\tilde{z}|\tilde{v}'} \var(\tilde{z},\tilde{z}'),
\end{aligned}
\end{equation}
respectively, where we denoted by $\tilde{\eth}'_{\scr{C}\tilde{z}|\tilde{v}'}$ the Weyl covariant derivative evaluated at the time $\tilde{v}'$. Putting the linear and the quadratic contributions together, we finally obtain the action of the non-local spin-1 charge on $\bar{\tilde{h}}$, which takes the following form
\begin{equation}
\begin{aligned}
\{\tilde{\cQ}_1(\tilde{v}, \tilde{z}), \bar{\tilde{h}}(\tilde{v}', \tilde{z}')\} &= - \varkappa^2\Bigl[ (\tilde{v} - \tilde{v}')\theta(\tilde{v}' - \tilde{v}) \tilde{\eth}^{'3}_{\scr{C}\tilde{z}}\var(\tilde{z},\tilde{z}') \\
&+\tilde{\eth}'_{\scr{C}\tilde{z}}\pa_{\tilde{v}'}[(\tilde{v} - \tilde{v}')\tilde{\lambda}(\tilde{v}', \tilde{z}) \theta(\tilde{v}' - \tilde{v})\var(\tilde{z},\tilde{z}')]\\
&-2\tilde{\lambda}(\tilde{v}',\tilde{z}) \theta(\tilde{v}' - \tilde{v}) \tilde{\eth}'_{\scr{C}\tilde{z}|\tilde{v}'}\var(\tilde{z},\tilde{z}')\Bigl].
\end{aligned}
\label{unr_qh}
\end{equation}
However, it is easy to note that the bracket diverges as $\tilde{v}\to -\infty$. In order to absorb these divergences, we define the following renormalized charge
\begin{equation}
\tilde{q}_1 = \f{1}{\varkappa^2}\Bigl(\tilde{\cQ}_1 - \tilde{v} \tilde{\eth}'_{\scr{C}}\tilde{\cQ}_0\Bigl),
\end{equation}
which is finite when $\tilde{v}\to -\infty$. Indeed, subtracting the contribution $\tilde{v} \tilde{\eth}'_{\scr{C}}\{\tilde{\cQ}_0(\tilde{v},\tilde{z}), \bar{\tilde{h}}(\tilde{v}',\tilde{z}')\}$ from \eqref{unr_qh}, the divergent terms cancel each other out and for $\tilde{v}'>\tilde{v}$ we obtain
\begin{equation}
\begin{aligned}
\{\tilde{q}_1(\tilde{z}), \bar{\tilde{h}}(\tilde{v}', \tilde{z}')\} &=  \tilde{v}'\tilde{\eth}'^3_{\scr{C}\tilde{z}} \var(\tilde{z}, \tilde{z}') +\tilde{\eth}'_{\scr{C}\tilde{z}} \pa_{\tilde{v}'} \Bigl[ \tilde{v}' \tilde{\lambda}(\tilde{v}',\tilde{z})\var(\tilde{z}, \tilde{z}') \Bigl]\\
&\quad + 2\tilde{\lambda}(\tilde{v}',\tilde{z})\tilde{\eth}'_{\scr{C}\tilde{z}|\tilde{v}'} \var(\tilde{z}, \tilde{z}').
\end{aligned}
\label{q1_h}
\end{equation}
For later convenience, it is useful to rewrite \eqref{q1_h} using the following Leibniz rule
\begin{equation}
f(\tilde{z}) \tilde{\eth}'^s_{\scr{C}\tilde{z}}\var(\tilde{z},\tilde{z}') = \sum_{n=0}^s(-1)^n\binom{s}{n} \tilde{\eth}'^n_{\scr{C}\tilde{z}'}f(\tilde{z}') \tilde{\eth}'^{s-n}_{\scr{C}\tilde{z}}\var(\tilde{z},\tilde{z}'),
\end{equation}
from which \eqref{q1_h} is re-arranged as follows
\begin{equation}
\begin{aligned}
\{\tilde{q}_1(\tilde{z}), \bar{\tilde{h}}(\tilde{v}', \tilde{z}')\} &= \tilde{v}'\tilde{\eth}'^3_{\scr{C}\tilde{z}} \var(\tilde{z}, \tilde{z}') + (\tilde{v}'\pa_{\tilde{v}'}+3)\tilde{\lambda}(\tilde{v}',\tilde{z})\tilde{\eth}'_{\scr{C}\tilde{z}} \var(\tilde{z}, \tilde{z}') \\
&\qquad -2 \tilde{\eth}'_{\scr{C}\tilde{z}'} \tilde{\lambda}(\tilde{v}',\tilde{z}') \var(\tilde{z}, \tilde{z}').
\end{aligned}
\label{q1_h_1}
\end{equation}
By introducing a parameter $\tilde{\tau}_1$, we can define the following near-horizon generator of the spin-1 symmetry
\begin{equation}
\tilde{Q}_1 = \int_{\tilde{\scr{S}}} \tilde{\tau}_1(z)\tilde{q}_1(\tilde{v},\tilde{z})\ \eps_{\tilde{\scr{S}}},
\end{equation}
whose charge aspect is $\tilde{q}_1$, and its action on $\bar{\tilde{h}}$ takes the following form
\begin{equation}
\begin{aligned}
\var_1 \bar{\tilde{h}} &:= \{\tilde{Q}_1, \bar{\tilde{h}}\} \\
&=-\tilde{v} \tilde{\eth}^{'3}_{\scr{C}\tilde{z}} \tilde{\tau}_1(\tilde{z}) -\tilde{\eth}'_{\scr{C}\tilde{z}}\Bigl[(\tilde{v}\pa_{\tilde{v}} +3)\tilde{\lambda}(\tilde{v}, \tilde{z}) \tilde{\tau}_1(\tilde{z})\Bigl] -2\tilde{\tau}_1(\tilde{z}) \tilde{\eth}'_{\scr{C}\tilde{z}} \tilde{\lambda}(\tilde{v}, \tilde{z}).
\end{aligned}
\end{equation}

\subsection{$\tilde{\cQ}_2$ charge action}
Finally, we want to evaluate the action of the non-local spin-2 charge $\tilde{\cQ}_{2}$ on $\bar{\tilde{h}}$. The charge expression can be organized into five contributions
\begin{equation}
\begin{aligned}
\tilde{\cQ}_{2}(\tilde{v}, \tilde{z})&= \tilde{\eth}^{'4}_{\scr{C}}\pa_{\tilde{v}}^{-3}\sigma -\tilde{\eth}^{'2}_{\scr{C}}\pa_{\tilde{v}}^{-3}(\tilde{\lambda} \pa_{\tilde{v}}\tilde{\sigma}) - 2\tilde{\eth}'_{\scr{C}} \pa_{\tilde{v}}^{-2}  (\tilde{\lambda} \tilde{\eth}'_{\scr{C}}\tilde{\sigma}) -3\pa_{\tilde{v}}^{-1}(\tilde{\lambda}\pa_{\tilde{v}}^{-1}\tilde{\eth}'^2_{\scr{C}}\tilde{\sigma}) \\
&\qquad +3\pa_{\tilde{v}}^{-1}(\tilde{\lambda}\pa_{\tilde{v}}^{-1}(\tilde{\lambda} \pa_{\tilde{v}}\tilde{\sigma})).
\end{aligned}
\label{q2_contr}
\end{equation}
The linear contribution follows straightforwardly and yields
\begin{equation}
\begin{aligned}
\{\tilde{\eth}'^4_{\scr{C}\tilde{z}}\pa_{\tilde{v}}^{-3} \tilde{\sigma}(\tilde{v}, \tilde{z}), \bar{\tilde{h}}(\tilde{v}', \tilde{z}')\} =-\f{\varkappa^2}{2}(\tilde{v}-\tilde{v}')^2 \theta (\tilde{v}'-\tilde{v})\tilde{\eth}'^4_{\scr{C}\tilde{z}}\var^{(2)}(\tilde{z}'-\tilde{z}).
\end{aligned}
\end{equation}
The next three terms represent the quadratic contributions to the spin-2 action and read
\begin{equation}
\begin{aligned}
\{ \tilde{\eth}'^2_{\scr{C}\tilde{z}}\pa_{\tilde{v}}^{-3} (\tilde{\lambda} \tilde{\cQ}_{-2})(\tilde{v}, \tilde{z}), \bar{\tilde{h}}(\tilde{v}', \tilde{z}')\} &= \varkappa^2 \tilde{\eth}'^2_{\scr{C}\tilde{z}}\pa_{\tilde{v}}^{-3} \Bigl(\tilde{\lambda}(\tilde{v}, \tilde{z}) \pa_{\tilde{v}} \var(\tilde{v} - \tilde{v}')\var(\tilde{z},\tilde{z}')\Bigl)\\
&= \f{\varkappa^2}{2} \tilde{\eth}'^2_{\scr{C}\tilde{z}}\pa_{\tilde{v}'} \Bigl[ (\tilde{v} - \tilde{v}')^2\tilde{\lambda}(\tilde{v}', \tilde{z})  \theta(\tilde{v}'-\tilde{v})\var(\tilde{z},\tilde{z}')\Bigl],
\end{aligned}
\end{equation}
then we have
\begin{equation}
\begin{aligned}
\{\tilde{\eth}'_{\scr{C}\tilde{z}} \pa_{\tilde{v}}^{-2} (\tilde{\lambda} \tilde{\cQ}_{-1})(\tilde{v}, \tilde{z}), \bar{\tilde{h}}(\tilde{v}', \tilde{z}')\} &= \tilde{\eth}'_{\scr{C}\tilde{z}} \pa_{\tilde{v}}^{-2}[\tilde{\lambda}(\tilde{v}, \tilde{z}) \{\tilde{\cQ}_{-1} (\tilde{v}, \tilde{z}), \bar{\tilde{h}}(\tilde{v}', \tilde{z}')\}]\\
&=\varkappa^2 \tilde{\eth}'_{\scr{C}\tilde{z}} \pa_{\tilde{v}}^{-2}[\tilde{\lambda}(\tilde{v}, \tilde{z})\var(\tilde{v}-\tilde{v}')\tilde{\eth}'_{\scr{C} \tilde{z}}\var(\tilde{z},\tilde{z}')]\\
&= -\varkappa^2 (\tilde{v}-\tilde{v}') \theta(\tilde{v}'-\tilde{v}) \tilde{\eth}'_{\scr{C}\tilde{z}}[ \tilde{\lambda}(\tilde{v}', \tilde{z}) \tilde{\eth}'_{\scr{C} \tilde{z}|\tilde{v}'} \var(\tilde{z},\tilde{z}')]\\
\end{aligned}
\end{equation}
and
\begin{equation}
\begin{aligned}
\{\pa_{\tilde{v}}^{-1} (\tilde{\lambda}\pa_{\tilde{v}}^{-1}\tilde{\eth}'_{\scr{C}\tilde{z}}\tilde{\cQ}_{-1})(\tilde{v}, \tilde{z}),  \bar{\tilde{h}}(\tilde{v}', \tilde{z}')\} &= \pa_{\tilde{v}}^{-1} [\tilde{\lambda}(\tilde{v}, \tilde{z})\pa_{\tilde{v}}^{-1}\tilde{\eth}'_{\scr{C}\tilde{z}}\{\tilde{\cQ}_{-1}(\tilde{v}, \tilde{z}),  \bar{\tilde{h}}(\tilde{v}', \tilde{z}')\}]\\
&=\varkappa^2\pa_{\tilde{v}}^{-1} [\tilde{\lambda}(\tilde{v}, \tilde{z})\pa_{\tilde{v}}^{-1}(\var(\tilde{v} - \tilde{v}')\tilde{\eth}'^2_{\scr{C}\tilde{z}}\var(\tilde{z},\tilde{z}'))]\\
&=-\varkappa^2\pa_{\tilde{v}}^{-1} [\tilde{\lambda}(\tilde{v}, \tilde{z}) \theta(\tilde{v}' - \tilde{v}) ]\tilde{\eth}'^2_{\scr{C}\tilde{z}|\tilde{v}'} \var(\tilde{z},\tilde{z}')\\
&=\varkappa^2 \Upsilon(\tilde{v}', \tilde{v}, \tilde{z})\theta(\tilde{v}' - \tilde{v}) \tilde{\eth}'^2_{\scr{C}\tilde{z}|\tilde{v}'} \var(\tilde{z},\tilde{z}'),
\end{aligned}
\end{equation}
respectively, where we introduced the history of the transversal shear $\Upsilon(\tilde{v}', \tilde{v}, \tilde{z})$, defined as follows
\begin{equation}
    \Upsilon (\tilde{v}', \tilde{v}, \tilde{z})  =\int_{\tilde{v}}^{\tilde{v}'}\dext {\tilde{v}''}\  \tilde{\lambda}(\tilde{v}'',\tilde{z}).
\end{equation} The last term in \eqref{q2_contr} is a cubic term and its contribution takes the following form
\begin{equation}
\begin{aligned}
\{\pa_{\tilde{v}}^{-1} (\tilde{\lambda}\pa_{\tilde{v}}^{-1} (\tilde{\lambda} \tilde{\cQ}_{-2}))(\tilde{v}, \tilde{z}),  \bar{\tilde{h}}(\tilde{v}', \tilde{z}')\} &= \pa_{\tilde{v}}^{-1} [\tilde{\lambda}(\tilde{v}, \tilde{z})\pa_{\tilde{v}}^{-1}(\tilde{\lambda}\{\tilde{\cQ}_{-2}(\tilde{v}, \tilde{z}),  \bar{\tilde{h}}(\tilde{v}', \tilde{z}')\})]\\
&=\varkappa^2 \pa_{\tilde{v}}^{-1} [\tilde{\lambda}(\tilde{v}, \tilde{z})\pa_{\tilde{v}}^{-1}(\tilde{\lambda} (\tilde{v}, \tilde{z}) \pa_{\tilde{v}}\var(\tilde{v}-\tilde{v}') \var(\tilde{z}, \tilde{z}')) ]\\
&=\varkappa^2\pa_{\tilde{v}}^{-1} \Bigl[\pa_{v'} \Bigl(\tilde{\lambda}(\tilde{v}, \tilde{z}) \tilde{\lambda} (\tilde{v}', \tilde{z})\theta(\tilde{v}' - \tilde{v})\var(\tilde{z},\tilde{z}') \Bigl)\Bigl]\\
&=-\varkappa^2\pa_{\tilde{v}'}\Bigl(\tilde{\lambda}(\tilde{v}', \tilde{z})\Upsilon(\tilde{v}', \tilde{v}, \tilde{z}) \theta(\tilde{v}' - \tilde{v})\var(\tilde{z},\tilde{z}')\Bigl ).
\end{aligned}
\end{equation}
The last two terms represent the non-local action of the $\tilde{\cQ}_2$ charge. Putting all the previous contributions together, we obtain
\begin{equation}
\begin{aligned}
\{\tilde{\cQ}_{2}(\tilde{v}, \tilde{z}), \bar{\tilde{h}}(\tilde{v}', \tilde{z}')\}&=-\f{\varkappa^2}{2}(\tilde{v}-\tilde{v}')^2 \theta (\tilde{v}'-\tilde{v})\tilde{\eth}_{\scr{C}\tilde{z}}'^4\var^{(2)}(\tilde{z}'-\tilde{z})\\
&\quad-\f{\varkappa^2}{2} \tilde{\eth}'^2_{\scr{C}\tilde{z}}\pa_{\tilde{v}'} \Bigl[ (\tilde{v} - \tilde{v}')^2\tilde{\lambda}(\tilde{v}', \tilde{z})  \theta(\tilde{v}'-\tilde{v})\var(\tilde{z},\tilde{z}')\Bigl]\\
&\quad +2\varkappa^2 (\tilde{v}-\tilde{v}') \theta(\tilde{v}'-\tilde{v}) \tilde{\eth}'_{\scr{C}\tilde{z}}[ \tilde{\lambda}(\tilde{v}', \tilde{z}) \tilde{\eth}'_{\scr{C} \tilde{z}|\tilde{v}'} \var(\tilde{z},\tilde{z}')]\\
&\quad -3\varkappa^2 \Upsilon(\tilde{v}', \tilde{v}, \tilde{z})\theta(\tilde{v}' - \tilde{v}) \tilde{\eth}'^2_{\scr{C}\tilde{z}|\tilde{v}'} \var(\tilde{z},\tilde{z}')\\
&\quad-3\varkappa^2\pa_{\tilde{v}'}\Bigl(\tilde{\lambda}(\tilde{v}', \tilde{z})\Upsilon(\tilde{v}', \tilde{v}, \tilde{z}) \theta(\tilde{v}' - \tilde{v})\var(\tilde{z},\tilde{z}')\Bigl ).
\end{aligned}
\label{Q_2h}
\end{equation}
Again, as $\tilde{v}\to -\infty$ the bracket \eqref{Q_2h} does not admit a well defined limit. Therefore, one has to renormalize the spin-2 charge by adding the following counter-terms
\begin{equation}
  \tilde{q}_2= \f{1}{\varkappa^2}\Bigl(\tilde{\cQ}_2  -\tilde{v}\tilde{\eth}'_{\scr{C}}\tilde{\cQ}_1 + \f{\tilde{v}^2}{2}\tilde{\eth}'^2_{\scr{C}}\tilde{\cQ}_0  - 3\tilde{\cQ}_0\pa^{-1}_{\tilde{v}} \tilde{\lambda}\Bigl),
\end{equation}
and the bracket becomes
\begin{equation}
\begin{aligned}
\{\tilde{q}_2(\tilde{z}), \bar{\tilde{h}}(\tilde{v}', \tilde{z}')\} &=-\f{\tilde{v}'^2}{2} \tilde{\eth}_{\scr{C}\tilde{z}}'^4\var^{(2)}(\tilde{z}'-\tilde{z})-\f{1}{2} \tilde{\eth}'^2_{\scr{C}\tilde{z}}\pa_{\tilde{v}'} \Bigl(\tilde{v}'^2\tilde{\lambda}(\tilde{v}', \tilde{z})  \var(\tilde{z},\tilde{z}')\Bigl)\\
&\quad -2 \tilde{v}' \tilde{\eth}'_{\scr{C}\tilde{z}}\Bigl( \tilde{\lambda}(\tilde{v}', \tilde{z}) \tilde{\eth}'_{\scr{C} \tilde{z}|\tilde{v}'} \var(\tilde{z},\tilde{z}')\Bigl) -3\Upsilon(\tilde{v}', +\infty, \tilde{z}) \tilde{\eth}'^2_{\scr{C}\tilde{z}|\tilde{v}'} \var(\tilde{z},\tilde{z}')\\
&\quad-3\pa_{\tilde{v}'}\Bigl(\tilde{\lambda}(\tilde{v}', \tilde{z})\Upsilon(\tilde{v}', +\infty, \tilde{z}) \var(\tilde{z},\tilde{z}')\Bigl ).
\end{aligned}
\label{q2_h}
\end{equation}
Similarly to the previous case, we can use the Leibniz rule to rewrite \eqref{q2_h} as follows
\begin{equation}
\begin{aligned}
\{\tilde{q}_2(\tilde{z}), \bar{\tilde{h}}(\tilde{v}', \tilde{z}')\} &=-\f{\tilde{v}'^2}{2} \tilde{\eth}_{\scr{C}\tilde{z}}'^4\var^{(2)}(\tilde{z}'-\tilde{z})\\
&\quad -\f{1}{2}\Bigl( (\tilde{v}'^2\pa_{\tilde{v}'} +6\tilde{v}'\pa_{\tilde{v}'} +6) \Upsilon(\tilde{v}',+\infty,\tilde{z})\Bigl)\tilde{\eth}'^2_{\scr{C},\tilde{z}}\var(\tilde{z},\tilde{z}')\\
&\quad +\f{1}{2}\Bigl( (\tilde{v}'\pa_{\tilde{v}'} +3)\tilde{\eth}'_{\scr{C}\tilde{z}} \Upsilon(\tilde{v}',+\infty,\tilde{z})\Bigl)\tilde{\eth}'_{\scr{C},\tilde{z}}\var(\tilde{z},\tilde{z}')\\
&\quad -3\var(\tilde{z},\tilde{z}')\tilde{\eth}'^2_{\scr{C}\tilde{z}'}\Upsilon(\tilde{v}',+\infty,\tilde{z}) -3 \pa_{\tilde{v}'}\Bigl(\tilde{\lambda}(\tilde{v}',\tilde{z})\Upsilon(\tilde{v}',+\infty,\tilde{z})\Bigl)\var(\tilde{z},\tilde{z}').
\end{aligned}
\label{141}
\end{equation}
Next, we introduce a parameter $\tilde{\tau}_2$ defined on $\scr{S}$, such that the generator of the spin-2 symmetry is 
\begin{equation}
\tilde{Q}_2 = \int_{\tilde{\scr{S}}} \tilde{\tau}_2(z)\tilde{q}_2(\tilde{v},\tilde{z})\ \eps_{\tilde{\scr{S}}},
\end{equation}
and it symmetry action on $\bar{\tilde{h}}$ assumes the following form
\begin{equation}
\begin{aligned}
\var_2 \bar{\tilde{h}} &:= \{\tilde{Q}_2, \bar{\tilde{h}}\} \\
&= -\f{\tilde{v}'}{2} \tilde{\eth}'^4_{\scr{C}}\tilde{\tau}_2 (\tilde{z}) -\f{1}{2}\tilde{\eth}'^2_{\scr{C}\tilde{z}}\Bigl[ (\tilde{v}^2\pa_{\tilde{v}} +6\tilde{v}\pa_{\tilde{v}} +6)\Upsilon(\tilde{v},+\infty,\tilde{z})\tilde{\tau}_2 (\tilde{z})\Bigl]\\
&\quad -\f{1}{2}\tilde{\eth}'_{\scr{C}\tilde{z}} \Bigl[(\tilde{v}\pa_{\tilde{v}} +3) \tilde{\eth}'_{\scr{C}\tilde{z}}\Upsilon(\tilde{v},+\infty,\tilde{z})\tilde{\tau}_2(\tilde{z}) \Bigl] -3\tilde{\tau}_2(\tilde{z}) \tilde{\eth}'^2_{\scr{C}\tilde{z}}\Upsilon(\tilde{v},+\infty,\tilde{z})\\
&\quad -3\tilde{\tau}_2(\tilde{z}) \pa_{\tilde{v}}\Bigl(\tilde{\lambda}(\tilde{v},\tilde{z}) \Upsilon(\tilde{v},+\infty,\tilde{z})\Bigl).
\end{aligned}
\end{equation}
This computation concludes the analysis of the action of the non-local charges $\tilde{\cQ}_s$ on the gravitational mode $\tilde{h}$, up to spin 2. The structure of the renormalized charges derived in these subsections, allows us to write down the expression of the non-local charges of arbitrary spin $s$. In the following section, we compute the linear and the quadratic action of the higher spin charges on $\tilde{h}$.

\subsection{Linear action}
Now that we computed the action and revealed the structure of renormalized non-local $\tilde{q}_s$ charges for $s\leq 2$, we can evaluate the general expression for the action of an arbitrary spin $s$ charges on $\tilde{h}$. The linear part of the renormalized spin-$s$ charge takes the following form 
\begin{equation}
\tilde{q}^1_s (\tilde{v},\tilde{z}) := \sum_{n=0}^{s} \frac{(-\tilde{v})^{n}}{n!} \tilde{\eth}'^{n}_{\scr{C}}\tilde{\cQ}_{s-n}^1 = \tilde{\eth}'^{s+2}_{\scr{C}}\sum_{n=0}^s \frac{(-\tilde{v})^{n}}{n!}  \pa^{n-s-1}_{\tilde{v}} \tilde{\sigma},
\label{ren_ch}
\end{equation}
where we used the fact that $\tilde{Q}^1_{s-n}= (\tilde{\eth}'_{\scr{C}}\pa^{-1}_{\tilde{v}})^{s-n+2} \pa_{\tilde{v}}\tilde{\sigma}$. Now, imposing the following condition
\begin{equation}
\lim_{\tilde{v}\to- \infty} \tilde{q}_s (\tilde{v},\tilde{z}) = \tilde{q}_s (\tilde{z})
\end{equation}
in \eqref{ren_ch},
and using the identity in \eqref{Leinb_2}, we obtain that
\begin{equation}
\lim_{\tilde{v}\to- \infty}\tilde{q}_s^1 (\tilde{v},\tilde{z}) = \lim_{\tilde{v}\to- \infty} \tilde{\eth}'^{s+2}_{\scr{C}}\pa_{\tilde{v}}^{-1}\Bigl(\f{\tilde{v}^s}{s!}\tilde{\sigma}(\tilde{v},\tilde{z})\Bigl) = \tilde{\eth}'^{s+2}_{\scr{C}} \tilde{\sigma}_s(\tilde{z}),
\label{lim_qs}
\end{equation}
where we defined the analogue of the (sub)$^s$-leading soft graviton operator in \cite{Freidel:2021dfs} for a generic null hypersurface,
\begin{equation}
\tilde{\sigma}_s(\tilde{z}):= \f{(-1)^s}{s!}\int_{-\infty}^{+\infty} \dext \tilde{v}\  \tilde{v}^s \tilde{\sigma}(\tilde{v},\tilde{z}).
\label{sigma_s}
\end{equation}
Next, our goal is to compute the linear action of $\tilde{q}_s$ on the gravitational mode $\bar{\tilde{h}}$. Using the expression in \eqref{ren_ch} and the asymptotic behaviour defined in \eqref{lim_qs}, we obtain
\begin{equation}
\begin{aligned}
\{\tilde{q}^1_s(\tilde{z}), \bar{\tilde{h}}(\tilde{v}',\tilde{z}')\} &=-\f{(-\tilde{v}')^s}{s!}\tilde{\eth}'^{s+2}_{\scr{C}}\var(\tilde{z},\tilde{z}').
\end{aligned}
\label{lin_act}
\end{equation}
It is straightforward to check that the formula \eqref{lin_act} reproduces the previously computed linear term for $s=0,1,2$.

\subsection{Quadratic action}
The quadratic contribution of the renormalized non-local charge of a generic spin $s$ reads as follows
\begin{equation}
\begin{aligned}
\varkappa^{2}\tilde{q}^2_s (\tilde{v},\tilde{z}) &= \sum_{n=0}^{s} \frac{(-\tilde{v})^{s-n}}{(s-n)!} \tilde{\eth}'^{s-n}_{\scr{C}}\tilde{\cQ}_{n}^2\\
&=  -\sum_{n=0}^{s} \sum_{\ell=0}^{n} \frac{(-\tilde{v})^{s-n} (\ell+1)}{(s-n)!} \tilde{\eth}'^{s-\ell}_{\scr{C}}  \pa_{\tilde{v}}^{-n+\ell-1} (\tilde{\lambda} \pa_{\tilde{v}}(\pa_{\tilde{v}}^{-1}\tilde{\eth}'_{\scr{C}})^\ell \tilde{\sigma}),
\end{aligned}
\end{equation}
where we used the formula \eqref{quad_ch} in the second line. Then, using the above formula we compute the quadratic action of a generic spin-$s$ charge on $\bar{\tilde{h}}$, which yields
\begin{equation}
\begin{aligned}
\{\tilde{q}^2_{s}(\tilde{v},\tilde{z}), \bar{\tilde{h}}(\tilde{v}',\tilde{z}')\}&= -\f{1}{\varkappa^2} \sum_{n=0}^{s}\sum_{\ell=0}^{n} \frac{(-\tilde{v})^{s-n}(\ell+1)}{(s-n)!} \tilde{\eth}'^{s-\ell}_{\scr{C}\tilde{z}}\pa_{\tilde{v}}^{\ell-n-1} \Bigl(\tilde{\lambda}\pa_{\tilde{v}}^{1-\ell}\tilde{\eth}'^\ell_{\scr{C}\tilde{z}} \{\tilde{\sigma}(\tilde{v},\tilde{z}), \bar{\tilde{h}}(\tilde{v}',\tilde{z}')\}\Bigl)\\
&= -\sum_{n=0}^{s}\sum_{\ell=0}^{n} \frac{(-\tilde{v})^{s-n}(\ell+1)}{(s-n)!} \tilde{\eth}'^{s-\ell}_{\scr{C} \tilde{z}}\pa_{\tilde{v}}^{\ell-n-1} \Bigl(\tilde{\lambda}(\tilde{v},\tilde{z})\tilde{\eth}'^\ell_{\scr{C} \tilde{z}|\tilde{v}'} \var(\tilde{z},\tilde{z}')\\
&\qquad\times \pa_{\tilde{v}}^{1-\ell} \var(\tilde{v}-\tilde{v}')\Bigl)\\
&= -\sum_{n=0}^{s}\sum_{\ell=0}^{n} \frac{(-1)^{\ell-1}(-\tilde{v})^{s-n}(\ell+1)}{(s-n)!} \tilde{\eth}'^{s-\ell}_{\scr{C} \tilde{z}}\pa_{\tilde{v}'}^{1-\ell} \Bigl(\tilde{\lambda}(\tilde{v}',\tilde{z})\pa_{\tilde{v}}^{\ell-n-1} \var(\tilde{v}-\tilde{v}')\Bigl)\\
&\qquad\times \tilde{\eth}'^\ell_{\scr{C}\tilde{z}|\tilde{v}'}\var(\tilde{z},\tilde{z}') \\
&= \sum_{n=0}^{s}\sum_{\ell=0}^{n} \frac{(-1)^{\ell-1}(-\tilde{v})^{s-n}(\ell+1)}{(s-n)!} \tilde{\eth}'^{s-\ell}_{\scr{C} \tilde{z}}\pa_{\tilde{v}'}^{1-\ell} \Bigl(\tilde{\lambda}(\tilde{v}',\tilde{z})\theta(\tilde{v}'-\tilde{v}) \\
&\qquad\times \f{(\tilde{v}-\tilde{v}')^{n-\ell}}{(n-\ell)!} \Bigl)\tilde{\eth}'^\ell_{\scr{C} \tilde{z}|\tilde{v}'} \var(\tilde{z},\tilde{z}'),
\end{aligned}
\end{equation}
where in the third line we used the identity
\begin{equation}
\pa_{\tilde{v}}^{-a}\Bigl(f(\tilde{v})\pa_{\tilde{v}}^{-b}[\var(\tilde{v}-\tilde{v}')g(\tilde{v})]\Bigl)= g(\tilde{v}')(-1)^b\pa_{\tilde{v}'}^{-b}\Bigl(f(\tilde{v}')\pa_{\tilde{v}}^{-a} \var(\tilde{v}-\tilde{v}')\Bigl) .
\end{equation}
Using the commutativity of the summation 
and the following identity
\begin{equation}
    \sum_{n=\ell}^s\f{(-\tilde{v})^{s-n}(\tilde{v}-\tilde{v}')^{n-\ell} }{(s-n)!(n-\ell)!} = \f{(-\tilde{v}')^{s-\ell}}{(s-\ell)!},
\end{equation}
the quadratic action then becomes
\begin{equation}
\begin{aligned}
\{\tilde{q}^2_{s}(\tilde{z}), \bar{\tilde{h}}(\tilde{v}',\tilde{z}')\}
&= -\sum_{\ell=0}^{s} \f{(-1)^{\ell}(\ell+1)}{(s-\ell)!} \tilde{\eth}'^{s-\ell}_{\scr{C}\tilde{z}} \Bigl[\pa_{\tilde{v}'}^{1-\ell}\Bigl(\tilde{\lambda}(\tilde{v}',\tilde{z})(-\tilde{v}')^{s-\ell}\Bigl) \tilde{\eth}'^\ell_{\scr{C} \tilde{z}|\tilde{v}'} \var(\tilde{z},\tilde{z}')\Bigl]\\
&=- \sum_{\ell=0}^{s} \f{(-1)^{\ell}(\ell+1)(\Delta-\ell)_{s-\ell}}{(s-\ell)!} \tilde{\eth}'^{s-\ell}_{\scr{C}\tilde{z}} \Bigl[ \pa_{\tilde{v}'}^{1-s} \tilde{\lambda}(\tilde{v}',\tilde{z}) \tilde{\eth}'^\ell_{\scr{C}\tilde{z}|\tilde{v}'} \var(\tilde{z},\tilde{z}')\Bigl],
\end{aligned}
\end{equation}
where we used the formula \eqref{Leinb_3gen} in the second line.
Finally, we use the following relation from the Leibniz rule, with $\Upsilon^{1-s} = \pa_{\tilde{v}'}^{1-s} \tilde{\lambda}$,
\begin{equation}
\Upsilon^{1-s}(\tilde{v}',\tilde{z})\tilde{\eth}'^\ell_{\scr{C}\tilde{z}} \var(\tilde{z},\tilde{z}') = \sum_{n=0}^{\ell} (-1)^n\binom{\ell}{n}\tilde{\eth}'^n_{\scr{C} \tilde{z}'} \Upsilon^{1-s}(\tilde{v}',\tilde{z}') \tilde{\eth}'^{\ell-n}_{\scr{C} \tilde{z}}\var(\tilde{z},\tilde{z}'),
\end{equation}
to write the bracket as follows
\begin{equation}
\begin{aligned}
\{\tilde{q}^2_{s}(\tilde{z}), \bar{\tilde{h}}(\tilde{v}',\tilde{z}')\}
&=- \sum_{\ell=0}^{s} \sum_{n=0}^{\ell} \f{(-1)^{\ell+n}(\ell+1)(\Delta-\ell)_{s-\ell}}{(s-\ell)!}\binom{\ell}{n}\tilde{\eth}'^n_{\scr{C} \tilde{z}'|\tilde{v}'} \Bigl(\pa_{\tilde{v}'}^{1-s} \tilde{\lambda} (\tilde{v}',\tilde{z}')\Bigl)\\
&\qquad \times \tilde{\eth}'^{s-n}_{\scr{C}\tilde{z}|\tilde{v}'}\var(\tilde{z},\tilde{z}').
\end{aligned}
\end{equation}
Switching the order of the sums and using the following identity
\begin{equation}
    \sum^s_{\ell=n} \f{(\ell+1)!(\Delta-\ell)_{s-\ell}}{(\ell-n)!(s-\ell)!}=\f{(n+1)!}{(s-n)!}(\Delta+2)_{s-n}
\end{equation}
proved in appendix E of \cite{Freidel:2021ytz}, we finally obtain 
\begin{equation}
\begin{aligned}
\{\tilde{q}^2_{s}(\tilde{z}), \bar{\tilde{h}}(\tilde{v},\tilde{z}')\}&= -\sum_{n=0}^s(-1)^{s+n} \f{(n+1)(\Delta+2)_{s-n}}{(s-n)!} \tilde{\eth}'^n_{\scr{C}\tilde{z}'} \Bigl( \pa_{\tilde{v}}^{1-s}\tilde{\lambda}(\tilde{v},\tilde{z}')\Bigl)\\
&\qquad \times \tilde{\eth}'^{s-n}_{\scr{C}\tilde{z}}\var(\tilde{z},\tilde{z}').
\end{aligned}
\label{gen_q2s}
\end{equation}
By substituting $s=0,1,2$, the quadratic terms in the formulas \eqref{117}, \eqref{q1_h_1} and \eqref{141} are recovered. 
Now, using the bracket in \eqref{gen_q2s} and the definition in \eqref{sigma_s}, the action of the quadratic charge on $\bar{\tilde{\sigma}}_{s'} (\tilde{z}')$ can be computed as follows
\begin{equation}
\begin{aligned}
\{\tilde{q}^2_s (\tilde{z}) , \bar{\tilde{\sigma}}_{s'} (\tilde{z}')\}  &= \f{1}{2} \f{(-1)^{s'}}{s'!} \int_{-\infty}^{+\infty} \dext \tilde{v} \ \tilde{v}^{s'} \{\tilde{q}^2_s(\tilde{z}),\bar{\tilde{\sigma}}(\tilde{v},\tilde{z}')\} \\
&= \f{1}{4} \f{(-1)^{s'}}{s'!} \int_{-\infty}^{+\infty} \dext \tilde{v} \ \tilde{v}^{s'} \pa_{\tilde{v}}\{\tilde{q}^2_s(\tilde{z}),\bar{\tilde{h}}(\tilde{v},\tilde{z}')\}\\
&= -\f{1}{4} \sum_{n=0}^s \f{(-1)^{s'+s+n} (n+1)}{s'!(s-n)!} \int_{-\infty}^{+\infty} \dext \tilde{v} \  \tilde{v}^{s'} \pa_{\tilde{v}}(\Delta+2)_{s-n} \\
&\qquad \times \tilde{\eth}'^n_{\scr{C}\tilde{z}'} \Bigl( \pa_{\tilde{v}}^{1-s}\tilde{\lambda}(\tilde{v},\tilde{z}')\Bigl) \tilde{\eth}'^{s-n}_{\scr{C}\tilde{z}}\var(\tilde{z},\tilde{z}')\\
&= -\f{1}{4} \sum_{n=0}^s \f{(-1)^{s'+s+n} (n+1)}{s'!(s-n)!} \int_{-\infty}^{+\infty} \dext \tilde{v} \  (\Delta+3-s')_{s-n}\tilde{v}^{s'} \pa_{\tilde{v}} \\
&\qquad \times \tilde{\eth}'^n_{\scr{C}\tilde{z}'} \Bigl( \pa_{\tilde{v}}^{1-s}\tilde{\lambda}(\tilde{v},\tilde{z}')\Bigl) \tilde{\eth}'^{s-n}_{\scr{C}\tilde{z}}\var(\tilde{z},\tilde{z}')\\
&= -\f{\tilde{\mu}_\infty}{4} \sum_{n=0}^s \f{(-1)^{s'+s+n} (n+1)}{s'!(s-n)!} \int_{-\infty}^{+\infty} \dext \tilde{v} \  (\Delta+3-s')_{s-n}\tilde{v}^{s'} \pa_{\tilde{v}} \\
&\qquad \times \tilde{\eth}'^n_{\scr{C}\tilde{z}'} \Bigl( \pa_{\tilde{v}}^{-s}\bar{\tilde{\sigma}}(\tilde{v},\tilde{z}')\Bigl) \tilde{\eth}'^{s-n}_{\scr{C}\tilde{z}}\var(\tilde{z},\tilde{z}'),
\end{aligned}
\label{q2_sig}
\end{equation}
where we used the following relations
\begin{equation}
\pa_{\tilde{v}}(\Delta+k)_{n}=(\Delta+k+1)_{n}\pa_{\tilde{v}}, \qquad \tilde{v}(\Delta+k)_{n}= (\Delta+k-1)_{n} \tilde{v},
\end{equation}
and in the last line we exploited the spin-coefficient equation for $\pa_{\tilde{v}}\tilde{\lambda}$ under the asymptotic conditions \eqref{fall_off_sigm} and \eqref{fall_off_rest}. Moreover, thanks to the asymptotic conditions \eqref{fall_off_sigm}-\eqref{fall_off_rest}, any analytic function of the operator $\Delta$ integrates to zero and, therefore, the \eqref{q2_sig} becomes
\begin{equation}
\begin{aligned}
\{\tilde{q}^2_s (\tilde{z}) , \bar{\tilde{\sigma}}_{s'} (\tilde{z}')\} 
&=-\f{\tilde{\mu}_\infty}{2} \sum_{n=0}^s \f{(n+1)(s+s'-n-4)_{s-n}}{(s-n)!} \tilde{\eth}'^n_{\scr{C}\tilde{z}'} \bar{\tilde{\sigma}}_{s+s'-1}(\tilde{z}') \tilde{\eth}'^{s-n}_{\scr{C}\tilde{z}}\var(\tilde{z},\tilde{z}')\\
&=-\f{\tilde{\mu}_\infty}{2} \sum_{n=0}^s (n+1)\binom{s+s'-n-4}{s'-4} \tilde{\eth}'^n_{\scr{C}\tilde{z}'} \bar{\tilde{\sigma}}_{s+s'-1}(\tilde{z}') \tilde{\eth}'^{s-n}_{\scr{C}\tilde{z}}\var(\tilde{z},\tilde{z}'),
\end{aligned}
\end{equation}
whose structure is the same as that found at null infinity \cite{Freidel:2021ytz, Geiller:2024bgf}. The commutator for the opposite spin is computed in appendix \ref{B} and reads
\begin{equation}
\{\tilde{q}^2_{s}(\tilde{v},\tilde{z}), \tilde{h}(\tilde{v}',\tilde{z}')\}=-\tilde{\mu}_\infty \sum_{\ell=0}^{s} \f{(\ell+1)(\Delta-2)_{s-\ell}}{(s-\ell)!} \pa_{\tilde{v}'}^{-s}\tilde{\eth}'^\ell_{\scr{C}\tilde{z}'} \tilde{\sigma}(\tilde{v}',\tilde{z}') \tilde{\eth}'^{s-\ell}_{\scr{C}\tilde{z}}\var(\tilde{z},\tilde{z}'),
\end{equation}
and following the same methodology used above, the action of the quadratic charge on the opposite spin $\tilde{\sigma}_{s'} (\tilde{z}')$ reads
\begin{equation}
\begin{aligned}
\{\tilde{q}^2_s (\tilde{z}) , \tilde{\sigma}_{s'} (\tilde{z}')\} 
&=-\f{\tilde{\mu}_\infty}{2} \sum_{n=0}^s (n+1)\binom{s+s'-n}{s'} \tilde{\eth}'^n_{\scr{C}\tilde{z}'} \tilde{\sigma}_{s+s'-1}(\tilde{z}') \tilde{\eth}'^{s-n}_{\scr{C}\tilde{z}}\var(\tilde{z},\tilde{z}').
\end{aligned}
\label{169}
\end{equation}

\subsection{Bracket at linear level}
To conclude, here we argue that the $w_{1+\infty}$ algebra follows straightforwardly by implementing exactly the same calculation performed in \cite{Freidel:2021ytz} or \cite{Geiller:2024bgf}. 
Indeed, the Poisson bracket at linear level is
\begin{equation}
\{\tilde{q}_s(\tilde{z}), \tilde{q}_{s'}(\tilde{z}')\}^1 = \{\tilde{q}^1_s(\tilde{z}), \tilde{q}^2_{s'}(\tilde{z}')\} + \{\tilde{q}^2_s(\tilde{z}), \tilde{q}^1_{s'}(\tilde{z}')\},
\end{equation}
and knowing that $\tilde{q}^1_s(\tilde{z})=\tilde{\eth}'^{s+2}_{\scr{C}} \tilde{\sigma}_s(\tilde{z})$, we have
\begin{equation}
\begin{aligned}
\{\tilde{q}^2_s(\tilde{z}), \tilde{q}^1_{s'}(\tilde{z}')\} &= \tilde{\eth}'^{s'+2}_{\scr{C}\tilde{z}'} \{\tilde{q}^2_s(\tilde{z}),  \tilde{\sigma}_{s'}(\tilde{z}')\}\\
&=-\f{\tilde{\mu}_\infty}{2} \sum_{n=0}^s (n+1)\binom{s+s'-n}{s'} \tilde{\eth}'^{s'+2}_{\scr{C}\tilde{z}'} \Bigl(\tilde{\eth}'^n_{\scr{C}\tilde{z}'} \tilde{\sigma}_{s+s'-1}(\tilde{z}') \tilde{\eth}'^{s-n}_{\scr{C}\tilde{z}}\var(\tilde{z},\tilde{z}')\Bigl).
\end{aligned}
\end{equation}
Since the expression above is completely independent of the dressing time $\tilde{v}$, the computations that follow are identical to that performed in appendix E of \cite{Freidel:2021ytz} or in appendix I of \cite{ Geiller:2024bgf}, with the substitution $\bar{N}_s\to\tilde{\sigma}_s$ and $\eth\to\tilde{\eth}'_{\scr{C}}$, and therefore the final result is
\begin{equation}
\begin{aligned}
\{\tilde{q}_s(\tilde{z}), \tilde{q}_{s'}(\tilde{z}')\}^1_{\tilde{\mu}_\infty} &= (s+1)\tilde{q}^1_{s'+s-1}(\tilde{z}) \tilde{\eth}'_{\scr{C}\tilde{z}'}\var(\tilde{z},\tilde{z}') \\
&\qquad -(s'+1)\tilde{q}^1_{s'+s-1}(\tilde{z}') \tilde{\eth}'_{\scr{C}\tilde{z}}\var(\tilde{z},\tilde{z}') ,
\end{aligned}
\end{equation}
where $\{\cdot,\cdot\}_{\tilde{\mu}_\infty} = \tilde{\mu}^{-1}_\infty\{\cdot,\cdot\}$, or
\begin{equation}
\{\tilde{Q}_{s_1}(\tilde{\tau}_1), \tilde{Q}_{s_2}(\tilde{\tau}_2)\}^1_{\tilde{\mu}_\infty} = (s_2+1)\tilde{Q}_{s_2+s_1-1}(\tilde{\tau}_2 \tilde{\eth}'_{\scr{C}} \tilde{\tau}_1) -(s_1+1)\tilde{Q}_{s_2+s_1-1}(\tilde{\tau}_1 \tilde{\eth}'_{\scr{C}}\tilde{\tau}_2)
\end{equation}
in terms of the higher spin charges.  In contrast to the asymptotic case, the presence of the cosmological constant does not complicate the calculations. Indeed, in the presence of a non-vanishing cosmological constant the asymptotic geometry is timelike or spacelike and the boundary metric is time dependent -- while $\scri^+$  behaves as an isolated horizon \cite{Ashtekar:2024bpi, Ashtekar:2024stm}. Therefore, it seems that the same procedure applied in this work to address the unfrozen degrees of freedom can be appropriately used for asymptotically (A)dS spacetimes. We leave this investigation for future work.

\section{Conclusions}
In this manuscript we derived the $w_{1+\infty}$ algebra for an arbitrary null hypersurface, working in what we have termed the Weyl reference frame — a dynamical reference frame in which the leading order of the longitudinal expansion $\varrho$ and the spin coefficient $\varepsilon$ are set to vanish through a combination of boost and Weyl rescaling. We emphasize that, while the Bianchi identities transform covariantly under Weyl rescalings, the null Raychaudhuri equation does not; consequently, the longitudinal shear is not constrained to vanish by the Raychaudhuri equation in this frame. Moreover, unlike the asymptotic flat case, further asymptotic conditions must be imposed on the boundary data $\tilde{\lambda}, \tilde{\mu}$ and $\tilde{\pi}$, in addition to those on the longitudinal shear $\tilde{\sigma}$. As argued in section \ref{sec3}, these fall-off conditions are necessary in order to recover the $w_{1+\infty}$ algebra at finite distance.\\
Working in the Weyl reference frame allows a direct integration of the Bianchi identities through the pseudo-differential operator $\partial_{\tilde{v}}^{-1}$, where $\tilde{v}$ is the Weyl dressing time, after which the methodology of \cite{Freidel:2021dfs, Freidel:2021ytz, Geiller:2024bgf} can be applied. We showed, however, that several features distinguish this construction from the asymptotic null case. In particular, the Weyl covariant derivative does not commute with the time-derivative operator, since the dressed corner metric depends on the dressing time $\tilde{v}$. This dependence produces non-local terms already at spin $s \leq 2$. To recover a bracket structure analogue to the asymptotic one, we absorbed these non-local terms into the definition of the charges; the resulting corrections admit a natural interpretation as longitudinal memory terms. Having obtained the general expression for the non-local charge of arbitrary spin $s$, we applied the machinery of \cite{Freidel:2021ytz, Geiller:2024bgf} to compute the bracket between two such charges at linear level, recovering an algebra of the $w_{1+\infty}$ type.
\\
Our results differ in scope and method from those of \cite{Ruzziconi:2025fuy}. There, the authors relate the subleading phase space at finite distance to the Ashtekar–Streubel phase space at null infinity by imposing self-duality conditions, effectively importing the asymptotic results to a generic null hypersurface once this identification is established. By contrast, in the present work we retain the genuine finite-distance degrees of freedom throughout, using them to construct a dynamical reference frame rather than mapping the system onto its asymptotic counterpart. We regard this distinction as physically significant, since it is precisely these unfrozen degrees of freedom that are responsible for the qualitative differences between the memory effect at $\scri^+$ and the memory effect on a black hole horizon \cite{Rahman:2019bmk}; we intend to return to this comparison in future work. \\
Throughout this work we have retained only the bulk contribution to the symplectic form, discarding the corner term $\Omega^c_{\pa\scr H}$; a complete treatment of the edge-mode content of the theory requires its inclusion, and we leave this analysis for future investigations. 
\\
Several directions for future work follow naturally from this construction. It would be interesting to extend the algebroid treatment of \cite{Cresto:2024fhd, Cresto:2024mne} to the finite-distance setting considered here, and to explore its relation to the twistor approach of \cite{Adamo:2021lrv, Kmec:2024nmu}. We further believe that the methods developed in this work can be applied to the analysis of the asymptotic higher-spin algebra in the presence of a non-vanishing cosmological constant; unlike the case treated here, the asymptotic geometry in that setting depends explicitly on the cosmological constant, and the definition of an appropriate news tensor is correspondingly less straightforward, so that this extension deserves separate and careful treatment \cite{Saw:2016isu, Mao:2019ahc, Compere:2019bua, Compere:2020lrt, Geiller:2022vto, Taylor:2023ajd}. Finally, it would be worthwhile to extend the present results to the Einstein–Maxwell theory, along the lines of \cite{DeSimone:2025ouu}.

\section*{Acknowledgments}
The author would like to thank Laurent Freidel and Daan Janssen for a fruitful interaction during  the Semi-Local Quantum Physics conference (SLQP 2026) in York, UK.

\appendix
\section{Weyl + class III transformations}\label{weyl+III}
In this appendix we list the transformation rules of the spin coefficients under Weyl and class III transformations. Under a Weyl rescaling, the spin coefficients transforms as follows
\begin{equation}
\begin{aligned}
\kappa&\to \mathbb{\Omega}^{3\omega_0-\omega_1}\kappa, \qquad &\varepsilon&\to \mathbb{\Omega}^{2\omega_0} [\varepsilon + (\omega_0+1)\mathbb{D}\ln\mathbb{\Omega}],\\
\nu &\to \mathbb{\Omega}^{3\omega_1-\omega_0}\nu, &\qquad \gamma&\to \mathbb{\Omega}^{2\omega_1} [\gamma - (\omega_1+1)\mathbb{\Delta}\ln\mathbb{\Omega}],\\ 
\sigma &\to \mathbb{\Omega}^{2\omega_0}\sigma, \qquad &\tau &\to \mathbb{\Omega}^{\omega_0+\omega_1}(\tau - \vardelta\ln\mathbb{\Omega}),\\ 
\lambda &\to \mathbb{\Omega}^{2\omega_1}\lambda, \qquad &\pi &\to \mathbb{\Omega}^{\omega_0+\omega_1}(\pi + \bar{\vardelta} \ln\mathbb{\Omega}),\\ 
\varrho &\to \mathbb{\Omega}^{2\omega_0}(\varrho -\mathbb{D}\ln\mathbb{\Omega}), \qquad &\alpha &\to \mathbb{\Omega}^{\omega_0+\omega_1}(\alpha +\omega_0 \bar{\vardelta}\ln\mathbb{\Omega}),\\ 
\mu &\to \mathbb{\Omega}^{2\omega_1}(\mu +\mathbb{\Delta}\ln\mathbb{\Omega}), \qquad &\beta &\to \mathbb{\Omega}^{\omega_0+\omega_1}[\beta +(\omega_0+1) \vardelta\ln\mathbb{\Omega}],\\ 
\end{aligned}
\end{equation}
while under class III Lorentz (spin+boost) transformation we have
\begin{equation}
\begin{aligned}
    \kappa &\to e^{2\lambda_L} e^{i\vartheta}\kappa, &\qquad &\sigma \to e^{\lambda_L}e^{2i\vartheta}\sigma,\\
    \alpha &\to e^{-i\vartheta}\Bigl(\alpha +\f{i}{2}\bar{\vardelta}\vartheta+\f{1}{2}\bar{\vardelta}\lambda_L\Bigl), &\qquad &\varrho\to e^{\lambda_L}\varrho,\\
    \beta &\to e^{i\vartheta}\Bigl(\beta +\f{i}{2}\vardelta\vartheta+\f{1}{2}\vardelta \lambda_L\Bigl), &\qquad &\tau\to e^{i\vartheta}\tau,\\
    \varepsilon&\to e^{\lambda_L}\Bigl(\varepsilon +\f{1}{2}\mathbb{D} \lambda_L +\f{i}{2}\mathbb{D}\vartheta\Bigl), &\qquad &\pi\to e^{-i\vartheta}\pi,\\
    \gamma &\to e^{-\lambda_L}\Bigl(\gamma +\f{1}{2}\mathbb{\Delta} \lambda_L +\f{i}{2}\mathbb{\Delta}\vartheta\Bigl), &\qquad &\mu \to e^{-\lambda_L}\mu,\\
    \lambda&\to e^{-\lambda_L} e^{-2i\vartheta}\lambda, &\qquad &\nu\to e^{-2\lambda_L} e^{-i\vartheta}\nu.
\end{aligned}
\end{equation}
As we argued throughout the manuscript, the null Raychaudhuri equation is not Weyl-covariant and changes as follows
\begin{equation}
\begin{aligned}
\tilde{\mathbb{D}}\tilde{\varrho} -\bar{\tilde{\vardelta}}\tilde{\kappa} &=\tilde{\varrho}^2 + \tilde{\sigma}\bar{\tilde{\sigma}} +(\tilde{\varepsilon} +\bar{\tilde{\varepsilon}})\tilde{\varrho} - \bar{\tilde{\kappa}}\tilde{\tau} - \tilde{\kappa}(3\tilde{\alpha}+\bar{\tilde{\beta}} -\tilde{\pi})+\mathbb{\Omega}^{4\omega_0} \Bigl[ (\mathbb{D}\ln\mathbb{\Omega})^2 \\
&\quad -\mathbb{D}^2\ln\mathbb{\Omega} +(\varepsilon+\bar{\varepsilon})\mathbb{D} \ln\mathbb{\Omega} -\bar{\kappa}\vardelta\ln\mathbb{\Omega} +(\omega_0+\omega_1) \kappa\bar{\vardelta}\ln\mathbb{\Omega} \Bigl],
\end{aligned}
\end{equation}
and also the equation for the transversal shear $\lambda$ becomes
\begin{equation}
\begin{aligned}
\tilde{\mathbb{D}}\tilde{\lambda} -\bar{\tilde{\vardelta}}\tilde{\pi}&= \tilde{\varrho}\tilde{\lambda} +\bar{\tilde{\sigma}}\tilde{\mu} +\tilde{\pi}^2 + (\tilde{\alpha} - \bar{\tilde{\beta}})\tilde{\pi} -\tilde{\nu}\bar{\tilde{\kappa}} -(3\tilde{\varepsilon} - \bar{\tilde{\varepsilon}}) \tilde{\lambda}+\mathbb{\Omega}^{2(\omega_0+\omega_1)} \Bigl[(2\omega_0+2\omega_1 \\
&\quad +3)\lambda\mathbb{D}\ln\mathbb{\Omega}  -(\omega_0+\omega_1+1)\pi\bar{\vardelta}\ln\mathbb{\Omega} -(\omega_0+\omega_1)(\bar{\vardelta}\ln \mathbb{\Omega})^2 -\bar{\vardelta}^2\ln\mathbb{\Omega}\\
&\qquad -\bar{\sigma}\mathbb{\Delta} \ln \mathbb{\Omega} -(\alpha-\bar{\beta}) \bar{\vardelta}\ln \mathbb{\Omega}\Bigl].
\end{aligned}
\end{equation}

\section{Commutator \eqref{long_mem}}\label{app_B}
In section \ref{Non-local charges}, we showed that in order to recover the same structure found at null infinity, non-local charges have to be defined via the commutator in \eqref{long_mem}. In particular, the commutator \eqref{long_mem} explicitly reads
\begin{equation}
\begin{aligned}
[\pa_{\tilde{v}}^{-1}, \tilde{\eth}'_{\scr{C}}] f(\tilde{v}, \tilde{z})&= \pa_{\tilde{v}}^{-1}\tilde{\eth}'_{\scr{C}} f(\tilde{v}, \tilde{z}) - \tilde{\eth}'_{\scr{C}}\pa_{\tilde{v}}^{-1} \it{f}(\tilde{v}, \tilde{z}) \\
&= \int_{+\infty}^{\tilde{v}}\dext {\tilde{v}'}\ \tilde{\eth}'_{\scr{C}} f(\tilde{v}', \tilde{z}) - \tilde{\eth}'_{\scr{C}}\int_{+\infty}^{\tilde{v}}\dext {\tilde{v}'}\ \it{f}(\tilde{v}', \tilde{z})\\
&= \int_{+\infty}^{\tilde{v}}\dext {\tilde{v}'}\ (\tilde{\eth}'_{\scr{C}|{\tilde{v}'}} - \tilde{\eth}'_{\scr{C}|{\tilde{v}}}) f(\tilde{v}', \tilde{z})\\
&= \int_{+\infty}^{\tilde{v}}\dext {\tilde{v}'}\ \Bigl(\int_{\tilde{v}}^{\tilde{v}'}\dext {\tilde{v}''}\ \pa_{\tilde{v}''}\tilde{\eth}'_{\scr{C}(\tilde{v}'')} \Bigl) f(\tilde{v}', \tilde{z})\\
&:= \int_{+\infty}^{\tilde{v}}\dext {\tilde{v}'}\ \Delta^{(f)}({\tilde{v}'}, \tilde{v} , \tilde{z}) \\
&= \pa^{-1}_{\tilde{v}}\Delta^{(f)}({\tilde{v}'},\tilde{v} , \tilde{z}).
\end{aligned}
\end{equation}
Using the above result, raising the derivative order of the Weyl covariant derivative, we obtain
\begin{equation}
\begin{aligned}
[\pa_{\tilde{v}}^{-1}, \tilde{\eth}'^2_{\scr{C}}] f(\tilde{v}, \tilde{z})&= \pa_{\tilde{v}}^{-1}\tilde{\eth}'^2_{\scr{C}} f(\tilde{v}, \tilde{z}) - \tilde{\eth}'^2_{\scr{C}}\pa_{\tilde{v}}^{-1} \it{f}(\tilde{v}, \tilde{z}) \\
&= [\pa_{\tilde{v}}^{-1},\tilde{\eth}'_{\scr{C}}] \tilde{\eth}'_{\scr{C}}f(\tilde{v}, \tilde{z})
- \tilde{\eth}'_{\scr{C}}[\tilde{\eth}'_{\scr{C}},\pa_{\tilde{v}}^{-1}] \it{f}(\tilde{v}, \tilde{z})\\
&= \pa^{-1}_{\tilde{v}}\Delta^{(\tilde{\eth}'_{\scr{C}}f)}({\tilde{v}'},\tilde{v} , \tilde{z}) 
+ \tilde{\eth}'_{\scr{C}} \pa^{-1}_{\tilde{v}}\Delta^{(f)}({\tilde{v}'},\tilde{v} , \tilde{z})\\
\end{aligned}
\end{equation}
and
\begin{equation}
\begin{aligned}
[\pa_{\tilde{v}}^{-1}, \tilde{\eth}'^3_{\scr{C}}] f(\tilde{v}, \tilde{z})&= \pa_{\tilde{v}}^{-1}\tilde{\eth}'^3_{\scr{C}} f(\tilde{v}, \tilde{z}) - \tilde{\eth}'^3_{\scr{C}}\pa_{\tilde{v}}^{-1} f(\tilde{v}, \tilde{z}) \\
&= [\pa_{\tilde{v}}^{-1},\tilde{\eth}'_{\scr{C}}] \tilde{\eth}'^2_{\scr{C}} f(\tilde{v}, \tilde{z}) +\tilde{\eth}'_{\scr{C}}[\pa_{\tilde{v}}^{-1}, \tilde{\eth}'^2_{\scr{C}}]f(\tilde{v}, \tilde{z}) \\
&= \pa^{-1}_{\tilde{v}}\Delta^{(\tilde{\eth}'^2_{\scr{C}} f )}(\tilde{v}',\tilde{v} , \tilde{z}) + \tilde{\eth}'_{\scr{C}} \pa^{-1}_{\tilde{v}} \Delta^{(\tilde{\eth}'_{\scr{C}}f )}({\tilde{v}}',\tilde{v} , \tilde{z})\\
&\quad+ \tilde{\eth}'^2_{\scr{C}} \pa^{-1}_{\tilde{v}}\Delta^{(f)}({\tilde{v}}',\tilde{v} , \tilde{z})
\end{aligned}
\end{equation}
In particular, from the above equations one can extract the following behaviour
\begin{equation}
[\pa_{\tilde{v}}^{-1}, \tilde{\eth}'^n_{\scr{C}}] f(\tilde{v}, \tilde{z}) = \sum_{k=0}^{n-1} \tilde{\eth}'^k_{\scr{C}} \pa_{\tilde{v}}^{-1} \Delta^{(\tilde{\eth}'^{n-1-k}_{\scr{C}}f)}(\tilde{v}',\tilde{v}, \tilde{z}).
\end{equation}

\section{$\tilde{h}$-bracket\label{B}}
Using the fundamental bracket in \eqref{fun_PB} and the relation $\pa_{\tilde{v}}\tilde{h}=2\tilde{\sigma}$, the bracket between $\tilde{h}$ and $\bar{\tilde{h}}$ reads as follows
\begin{equation}
\{\tilde{h}(\tilde{v}, \tilde{z}), \tilde{\bar{h}}(\tilde{v}', \tilde{z}')\} = -2\varkappa^2 \theta(\tilde{v}'-\tilde{v})\var(\tilde{z},\tilde{z}'),
\end{equation}
and is useful in computing the bracket $\{\tilde{q}^2_{s}(\tilde{v},\tilde{z}), \tilde{h}(\tilde{v}',\tilde{z}')\}$. Indeed, we obtain for $\tilde{v}'>\tilde{v}$
\begin{equation}
\begin{aligned}
\{\tilde{q}^2_{s}(\tilde{v},\tilde{z}), \tilde{h}(\tilde{v}',\tilde{z}')\}&=  -\varkappa^{-2}\sum_{n=0}^{s} \sum_{\ell=0}^{n} \frac{(-\tilde{v})^{s-n} (\ell+1)}{(s-n)!} \tilde{\eth}'^{s-\ell}_{\scr{C}\tilde{z}}  \pa_{\tilde{v}}^{-n+\ell-1} \Bigl(\{\tilde{\lambda}(\tilde{v},\tilde{z}), \tilde{h}(\tilde{v}',\tilde{z}')\}\\
&\qquad\times \pa_{\tilde{v}}(\pa_{\tilde{v}}^{-1}\tilde{\eth}'_{\scr{C}\tilde{z}})^\ell \tilde{\sigma}\Bigl)\\
&\simeq \tilde{\mu}_\infty\sum_{n=0}^{s} \sum_{\ell=0}^{n} \frac{(-\tilde{v})^{s-n} (\ell+1)}{(s-n)!} \tilde{\eth}'^{s-\ell}_{\scr{C}\tilde{z}}\Bigl[\var(\tilde{z},\tilde{z}')\pa_{\tilde{v}}^{-n+\ell-1} \Bigl(\theta(\tilde{v}'-\tilde{v})\\
&\qquad\times \pa_{\tilde{v}}(\pa_{\tilde{v}}^{-1}\tilde{\eth}'_{\scr{C}\tilde{z}})^\ell \tilde{\sigma}\Bigl)\Bigl]\\
&=\tilde{\mu}_\infty\sum_{n=0}^{s} \sum_{\ell=0}^{n} \frac{(-\tilde{v})^{s-n} (\ell+1)}{(s-n)!} \tilde{\eth}'^{s-\ell}_{\scr{C}\tilde{z}} \Bigl[\var(\tilde{z},\tilde{z}')\pa_{\tilde{v}}^{-n+\ell-1} \pa_{\tilde{v}'}^{-1} \Bigl(\var(\tilde{v}'-\tilde{v})\\
&\qquad\times \pa_{\tilde{v}'}^{1-\ell}\tilde{\eth}'^\ell_{\scr{C}\tilde{z}} \tilde{\sigma}(\tilde{v}',\tilde{z})\Bigl)\Bigl]\\
&=-\tilde{\mu}_\infty\sum_{n=0}^{s} \sum_{\ell=0}^{n} \frac{(-\tilde{v})^{s-n} (\ell+1)}{(s-n)!} \tilde{\eth}'^{s-\ell}_{\scr{C}\tilde{z}}\Bigl[\var(\tilde{z},\tilde{z}')\pa_{\tilde{v}'}^{-1} \Bigl(\f{(\tilde{v} - \tilde{v}')^{n-\ell}}{(n-\ell)!}\\
&\qquad\times \theta(\tilde{v}'-\tilde{v}) \pa_{\tilde{v}'}^{1-\ell}\tilde{\eth}'^\ell_{\scr{C}\tilde{z}} \tilde{\sigma}(\tilde{v}',\tilde{z})\Bigl)\Bigl]\\
\end{aligned}
\end{equation}
where in the second line we used the fall-off conditions in \eqref{fall_off_rest}. By evaluating the sum $\sum_{n=\ell}^s$ first, sending $\tilde{v}\to-\infty$ and subsequently  using the identity in \eqref{Leinb_3gen}, we obtain 
\begin{equation}
\begin{aligned}
\{\tilde{q}^2_{s}(\tilde{v},\tilde{z}), \tilde{h}(\tilde{v}',\tilde{z}')\}&=-\tilde{\mu}_\infty \sum_{\ell=0}^{s} (\ell+1)\tilde{\eth}'^{s-\ell}_{\scr{C}\tilde{z}}\var(\tilde{z},\tilde{z}')\pa_{\tilde{v}'}^{-1} \Bigl(\f{(-\tilde{v}')^{s-\ell}}{(s-\ell)!}\pa_{\tilde{v}'}^{1-\ell}\tilde{\eth}'^\ell_{\scr{C}\tilde{z}'} \tilde{\sigma}(\tilde{v}',\tilde{z}')\Bigl) \\
&=-\tilde{\mu}_\infty \sum_{\ell=0}^{s} \f{(\ell+1)(\Delta-2)_{s-\ell}}{(s-\ell)!} \pa_{\tilde{v}'}^{-s}\tilde{\eth}'^\ell_{\scr{C}\tilde{z}'} \tilde{\sigma}(\tilde{v}',\tilde{z}') \tilde{\eth}'^{s-\ell}_{\scr{C}\tilde{z}}\var(\tilde{z},\tilde{z}').
\end{aligned}
\end{equation}
Then, the computation of the bracket
\begin{equation}
\begin{aligned}
\{\tilde{q}^2_{s}(\tilde{z}), \tilde{\sigma}_{s'}(\tilde{z}')\}&= \f{1}{2}\f{(-1)^{s'}}{s'!}\int_{-\infty}^{+\infty}\dd\tilde{v}\ \tilde{v}^{s'}\{\tilde{q}^2_s(\tilde{z}), \tilde{\sigma}(\tilde{v},\tilde{z}')\}\\
&=- \f{\tilde{\mu}_\infty}{4}\sum_{\ell=0}^{s} \f{(-1)^{s'}(\ell+1)}{(s-\ell)!s'!}\int_{-\infty}^{+\infty}\dd\tilde{v}\ \tilde{v}^{s'} \pa_{\tilde{v}}(\Delta-2)_{s-\ell}\\
&\qquad \times \pa_{\tilde{v}}^{-s}\tilde{\eth}'^\ell_{\scr{C}\tilde{z}'} \tilde{\sigma}(\tilde{v},\tilde{z}') \tilde{\eth}'^{s-\ell}_{\scr{C}\tilde{z}}\var(\tilde{z},\tilde{z}')
\end{aligned}
\end{equation}
is straightforward, since it follows the same methodology applied in \eqref{q2_sig}, and gives
\begin{equation}
\begin{aligned}
\{\tilde{q}^2_s (\tilde{z}) , \tilde{\sigma}_{s'} (\tilde{z}')\} 
&=-\f{\tilde{\mu}_\infty}{2} \sum_{n=0}^s (n+1)\binom{s+s'-n}{s'} \tilde{\eth}'^n_{\scr{C}\tilde{z}'} \tilde{\sigma}_{s+s'-1}(\tilde{z}') \tilde{\eth}'^{s-n}_{\scr{C}\tilde{z}}\var(\tilde{z},\tilde{z}').
\end{aligned}
\end{equation}

\bibliographystyle{unsrt}
\bibliography{bib}

@article{Ashtekar:2024bpi,
    author = "Ashtekar, Abhay and Speziale, Simone",
    title = "{Null infinity as a weakly isolated horizon}",
    eprint = "2402.17977",
    archivePrefix = "arXiv",
    primaryClass = "hep-th",
    doi = "10.1103/PhysRevD.110.044048",
    journal = "Phys. Rev. D",
    volume = "110",
    number = "4",
    pages = "044048",
    year = "2024"
}

@article{Ashtekar:2024stm,
    author = "Ashtekar, Abhay and Speziale, Simone",
    title = "{Null infinity and horizons: A new approach to fluxes and charges}",
    eprint = "2407.03254",
    archivePrefix = "arXiv",
    primaryClass = "hep-th",
    doi = "10.1103/PhysRevD.110.044049",
    journal = "Phys. Rev. D",
    volume = "110",
    number = "4",
    pages = "044049",
    year = "2024"
}

@article{Freidel:2021fxf,
    author = "Freidel, Laurent and Oliveri, Roberto and Pranzetti, Daniele and Speziale, Simone",
    title = "{The Weyl BMS group and Einstein{\textquoteright}s equations}",
    eprint = "2104.05793",
    archivePrefix = "arXiv",
    primaryClass = "hep-th",
    doi = "10.1007/JHEP07(2021)170",
    journal = "JHEP",
    volume = "07",
    pages = "170",
    year = "2021"
}

@article{Bondi:1962px,
    author = "Bondi, H. and van der Burg, M. G. J. and Metzner, A. W. K.",
    title = "{Gravitational waves in general relativity. 7. Waves from axisymmetric isolated systems}",
    doi = "10.1098/rspa.1962.0161",
    journal = "Proc. Roy. Soc. Lond. A",
    volume = "269",
    pages = "21--52",
    year = "1962"
}

@article{Sachs:1962wk,
    author = "Sachs, R. K.",
    title = "{Gravitational waves in general relativity. 8. Waves in asymptotically flat space-times}",
    doi = "10.1098/rspa.1962.0206",
    journal = "Proc. Roy. Soc. Lond. A",
    volume = "270",
    pages = "103--126",
    year = "1962"
}

@article{Barnich:2009se,
    author = "Barnich, Glenn and Troessaert, Cedric",
    title = "{Symmetries of asymptotically flat 4 dimensional spacetimes at null infinity revisited}",
    eprint = "0909.2617",
    archivePrefix = "arXiv",
    primaryClass = "gr-qc",
    reportNumber = "ULB-TH-09-24",
    doi = "10.1103/PhysRevLett.105.111103",
    journal = "Phys. Rev. Lett.",
    volume = "105",
    pages = "111103",
    year = "2010"
}

@article{Guevara:2021abz,
    author = "Guevara, Alfredo and Himwich, Elizabeth and Pate, Monica and Strominger, Andrew",
    title = "{Holographic symmetry algebras for gauge theory and gravity}",
    eprint = "2103.03961",
    archivePrefix = "arXiv",
    primaryClass = "hep-th",
    doi = "10.1007/JHEP11(2021)152",
    journal = "JHEP",
    volume = "11",
    pages = "152",
    year = "2021"
}

@article{Strominger:2021mtt,
    author = "Strominger, Andrew",
    title = "{$w_{1+\infty}$ Algebra and the Celestial Sphere: Infinite Towers of Soft Graviton, Photon, and Gluon Symmetries}",
    eprint = "2105.14346",
    archivePrefix = "arXiv",
    primaryClass = "hep-th",
    doi = "10.1103/PhysRevLett.127.221601",
    journal = "Phys. Rev. Lett.",
    volume = "127",
    number = "22",
    pages = "221601",
    year = "2021"
}

@article{Pasterski:2016qvg,
    author = "Pasterski, Sabrina and Shao, Shu-Heng and Strominger, Andrew",
    title = "{Flat Space Amplitudes and Conformal Symmetry of the Celestial Sphere}",
    eprint = "1701.00049",
    archivePrefix = "arXiv",
    primaryClass = "hep-th",
    doi = "10.1103/PhysRevD.96.065026",
    journal = "Phys. Rev. D",
    volume = "96",
    number = "6",
    pages = "065026",
    year = "2017"
}

@article{Pasterski:2017kqt,
    author = "Pasterski, Sabrina and Shao, Shu-Heng",
    title = "{Conformal basis for flat space amplitudes}",
    eprint = "1705.01027",
    archivePrefix = "arXiv",
    primaryClass = "hep-th",
    doi = "10.1103/PhysRevD.96.065022",
    journal = "Phys. Rev. D",
    volume = "96",
    number = "6",
    pages = "065022",
    year = "2017"
}

@article{Cheung:2016iub,
    author = "Cheung, Clifford and de la Fuente, Anton and Sundrum, Raman",
    title = "{4D scattering amplitudes and asymptotic symmetries from 2D CFT}",
    eprint = "1609.00732",
    archivePrefix = "arXiv",
    primaryClass = "hep-th",
    reportNumber = "CALT-TH-2016-024, UMD-PP-017-010",
    doi = "10.1007/JHEP01(2017)112",
    journal = "JHEP",
    volume = "01",
    pages = "112",
    year = "2017"
}

@book{Strominger:2017zoo,
    author = "Strominger, Andrew",
    title = "{Lectures on the Infrared Structure of Gravity and Gauge Theory}",
    eprint = "1703.05448",
    archivePrefix = "arXiv",
    primaryClass = "hep-th",
    isbn = "978-0-691-17973-5",
    month = "3",
    year = "2017"
}

@article{Raclariu:2021zjz,
    author = "Raclariu, Ana-Maria",
    title = "{Lectures on Celestial Holography}",
    eprint = "2107.02075",
    archivePrefix = "arXiv",
    primaryClass = "hep-th",
    month = "7",
    year = "2021"
}

@article{Pasterski:2021rjz,
    author = "Pasterski, Sabrina",
    title = "{Lectures on celestial amplitudes}",
    eprint = "2108.04801",
    archivePrefix = "arXiv",
    primaryClass = "hep-th",
    doi = "10.1140/epjc/s10052-021-09846-7",
    journal = "Eur. Phys. J. C",
    volume = "81",
    number = "12",
    pages = "1062",
    year = "2021"
}

@article{Strominger:2013jfa,
    author = "Strominger, Andrew",
    title = "{On BMS Invariance of Gravitational Scattering}",
    eprint = "1312.2229",
    archivePrefix = "arXiv",
    primaryClass = "hep-th",
    doi = "10.1007/JHEP07(2014)152",
    journal = "JHEP",
    volume = "07",
    pages = "152",
    year = "2014"
}

@article{He:2014laa,
    author = "He, Temple and Lysov, Vyacheslav and Mitra, Prahar and Strominger, Andrew",
    title = "{BMS supertranslations and Weinberg{\textquoteright}s soft graviton theorem}",
    eprint = "1401.7026",
    archivePrefix = "arXiv",
    primaryClass = "hep-th",
    doi = "10.1007/JHEP05(2015)151",
    journal = "JHEP",
    volume = "05",
    pages = "151",
    year = "2015"
}

@article{Gomes:2016mwl,
    author = "Gomes, Henrique and Riello, Aldo",
    title = "{The observer{\textquoteright}s ghost: notes on a field space connection}",
    eprint = "1608.08226",
    archivePrefix = "arXiv",
    primaryClass = "math-ph",
    doi = "10.1007/JHEP05(2017)017",
    journal = "JHEP",
    volume = "05",
    pages = "017",
    year = "2017"
}

@article{Donnelly:2016auv,
    author = "Donnelly, William and Freidel, Laurent",
    title = "{Local subsystems in gauge theory and gravity}",
    eprint = "1601.04744",
    archivePrefix = "arXiv",
    primaryClass = "hep-th",
    doi = "10.1007/JHEP09(2016)102",
    journal = "JHEP",
    volume = "09",
    pages = "102",
    year = "2016"
}

@article{Freidel:2021dxw,
    author = "Freidel, Laurent",
    title = "{A canonical bracket for open gravitational system}",
    eprint = "2111.14747",
    archivePrefix = "arXiv",
    primaryClass = "hep-th",
    month = "11",
    year = "2021"
}

@article{Speranza:2017gxd,
    author = "Speranza, Antony J.",
    title = "{Local phase space and edge modes for diffeomorphism-invariant theories}",
    eprint = "1706.05061",
    archivePrefix = "arXiv",
    primaryClass = "hep-th",
    doi = "10.1007/JHEP02(2018)021",
    journal = "JHEP",
    volume = "02",
    pages = "021",
    year = "2018"
}

@article{Ciambelli:2021nmv,
    author = "Ciambelli, Luca and Leigh, Robert G. and Pai, Pin-Chun",
    title = "{Embeddings and Integrable Charges for Extended Corner Symmetry}",
    eprint = "2111.13181",
    archivePrefix = "arXiv",
    primaryClass = "hep-th",
    doi = "10.1103/PhysRevLett.128.171302",
    journal = "Phys. Rev. Lett.",
    volume = "128",
    year = "2022"
}

@article{Carrozza:2021gju,
    author = "Carrozza, Sylvain and Hoehn, Philipp A.",
    title = "{Edge modes as reference frames and boundary actions from post-selection}",
    eprint = "2109.06184",
    archivePrefix = "arXiv",
    primaryClass = "hep-th",
    doi = "10.1007/JHEP02(2022)172",
    journal = "JHEP",
    volume = "02",
    pages = "172",
    year = "2022"
}

@article{Carrozza:2022xut,
    author = "Carrozza, Sylvain and Eccles, Stefan and Hoehn, Philipp A.",
    title = "{Edge modes as dynamical frames: charges from post-selection in generally covariant theories}",
    eprint = "2205.00913",
    archivePrefix = "arXiv",
    primaryClass = "hep-th",
    doi = "10.21468/SciPostPhys.17.2.048",
    journal = "SciPost Phys.",
    volume = "17",
    number = "2",
    pages = "048",
    year = "2024"
}

@article{Goeller:2022rsx,
    author = "Goeller, Christophe and Hoehn, Philipp A. and Kirklin, Josh",
    title = "{Diffeomorphism-invariant observables and dynamical frames in gravity: reconciling bulk locality with general covariance}",
    eprint = "2206.01193",
    archivePrefix = "arXiv",
    primaryClass = "hep-th",
    month = "6",
    year = "2022"
}

@article{Ciambelli:2023mir,
    author = "Ciambelli, Luca and Freidel, Laurent and Leigh, Robert G.",
    title = "{Null Raychaudhuri: canonical structure and the dressing time}",
    eprint = "2309.03932",
    archivePrefix = "arXiv",
    primaryClass = "hep-th",
    doi = "10.1007/JHEP01(2024)166",
    journal = "JHEP",
    volume = "01",
    pages = "166",
    year = "2024"
}

@article{Ciambelli:2024swv,
    author = "Ciambelli, Luca and Freidel, Laurent and Leigh, Robert G.",
    title = "{Quantum null geometry and gravity}",
    eprint = "2407.11132",
    archivePrefix = "arXiv",
    primaryClass = "hep-th",
    doi = "10.1007/JHEP12(2024)028",
    journal = "JHEP",
    volume = "12",
    pages = "028",
    year = "2024"
}

@article{Geiller:2022vto,
    author = "Geiller, Marc and Zwikel, C{\'e}line",
    title = "{The partial Bondi gauge: Further enlarging the asymptotic structure of gravity}",
    eprint = "2205.11401",
    archivePrefix = "arXiv",
    primaryClass = "hep-th",
    doi = "10.21468/SciPostPhys.13.5.108",
    journal = "SciPost Phys.",
    volume = "13",
    pages = "108",
    year = "2022"
}

@article{Freidel:2025ous,
    author = "Freidel, Laurent and Kirklin, Josh",
    title = "{Localization and anomalous reference frames in gravity}",
    eprint = "2510.26589",
    archivePrefix = "arXiv",
    primaryClass = "hep-th",
    month = "10",
    year = "2025"
}

@article{Ruzziconi:2025fuy,
    author = "Ruzziconi, Romain and Zwikel, C{\'e}line",
    title = "{Celestial $Lw_{1+\infty}$ symmetries and subleading phase space of null hypersurfaces}",
    eprint = "2511.07525",
    archivePrefix = "arXiv",
    primaryClass = "hep-th",
    doi = "10.1103/hrbd-cmr7",
    journal = "Phys. Rev. D",
    volume = "113",
    number = "4",
    pages = "044067",
    year = "2026"
}

@article{DeSimone:2026nig,
    author = "De Simone, Gianfranco",
    title = "{Duality symmetry and dynamics on finite null boundaries}",
    eprint = "2602.21017",
    archivePrefix = "arXiv",
    primaryClass = "hep-th",
    journal = "JHEP",
    volume = "07",
    pages = "076",
    year = "2026"
}

@article{Freidel:2021dfs,
    author = "Freidel, Laurent and Pranzetti, Daniele and Raclariu, Ana-Maria",
    title = "{Sub-subleading soft graviton theorem from asymptotic Einstein{\textquoteright}s equations}",
    eprint = "2111.15607",
    archivePrefix = "arXiv",
    primaryClass = "hep-th",
    doi = "10.1007/JHEP05(2022)186",
    journal = "JHEP",
    volume = "05",
    pages = "186",
    year = "2022"
}

@article{Freidel:2021ytz,
    author = "Freidel, Laurent and Pranzetti, Daniele and Raclariu, Ana-Maria",
    title = "{Higher spin dynamics in gravity and $w_{1+\infty}$ celestial symmetries}",
    eprint = "2112.15573",
    archivePrefix = "arXiv",
    primaryClass = "hep-th",
    doi = "10.1103/PhysRevD.106.086013",
    journal = "Phys. Rev. D",
    volume = "106",
    number = "8",
    pages = "086013",
    year = "2022"
}

@article{Geiller:2024bgf,
    author = "Geiller, Marc",
    title = "{Celestial $w_{1+\infty}$ charges and the subleading structure of asymptotically-flat spacetimes}",
    eprint = "2403.05195",
    archivePrefix = "arXiv",
    primaryClass = "hep-th",
    doi = "10.21468/SciPostPhys.18.1.023",
    journal = "SciPost Phys.",
    volume = "18",
    number = "1",
    pages = "023",
    year = "2025"
}

@article{DeSimone:2025ouu,
    author = "De Simone, Gianfranco",
    title = "{Near-Horizon Symmetries in Einstein-Maxwell theory}",
    eprint = "2511.11136",
    archivePrefix = "arXiv",
    primaryClass = "hep-th",
    doi = "10.1088/1402-4896/ae858a",
    journal = "Physica Scripta",
    month = "11",
    year = "2025"
}

@article{Rahman:2019bmk,
    author = "Rahman, Adel A. and Wald, Robert M.",
    title = "{Black Hole Memory}",
    eprint = "1912.12806",
    archivePrefix = "arXiv",
    primaryClass = "gr-qc",
    doi = "10.1103/PhysRevD.101.124010",
    journal = "Phys. Rev. D",
    volume = "101",
    number = "12",
    pages = "124010",
    year = "2020"
}

@article{Cresto:2024fhd,
    author = "Cresto, Nicolas and Freidel, Laurent",
    title = "{Asymptotic higher spin symmetries I: covariant wedge algebra in gravity}",
    eprint = "2409.12178",
    archivePrefix = "arXiv",
    primaryClass = "hep-th",
    doi = "10.1007/s11005-025-01921-4",
    journal = "Lett. Math. Phys.",
    volume = "115",
    number = "2",
    pages = "39",
    year = "2025"
}

@article{Cresto:2024mne,
    author = "Cresto, Nicolas and Freidel, Laurent",
    title = "{Asymptotic higher spin symmetries II: Noether realization in gravity}",
    eprint = "2410.15219",
    archivePrefix = "arXiv",
    primaryClass = "hep-th",
    doi = "10.1007/JHEP03(2026)147",
    journal = "JHEP",
    volume = "03",
    pages = "147",
    year = "2026"
}

@article{Ashtekar:1981bq,
    author = "Ashtekar, A. and Streubel, M.",
    title = "{Symplectic Geometry of Radiative Modes and Conserved Quantities at Null Infinity}",
    doi = "10.1098/rspa.1981.0109",
    journal = "Proc. Roy. Soc. Lond. A",
    volume = "376",
    pages = "585--607",
    year = "1981"
}

@article{Adamo:2021lrv,
    author = "Adamo, Tim and Mason, Lionel and Sharma, Atul",
    title = "{Celestial $w_{1+\infty}$ Symmetries from Twistor Space}",
    eprint = "2110.06066",
    archivePrefix = "arXiv",
    primaryClass = "hep-th",
    doi = "10.3842/SIGMA.2022.016",
    journal = "SIGMA",
    volume = "18",
    pages = "016",
    year = "2022"
}

@article{Kmec:2024nmu,
    author = "Kmec, Adam and Mason, Lionel and Ruzziconi, Romain and Yelleshpur Srikant, Akshay",
    title = "{Celestial $Lw_{1+\infty}$ charges from a twistor action}",
    eprint = "2407.04028",
    archivePrefix = "arXiv",
    primaryClass = "hep-th",
    doi = "10.1007/JHEP10(2024)250",
    journal = "JHEP",
    volume = "10",
    pages = "250",
    year = "2024"
}

@article{Geroch:1968zm,
    author = "Geroch, Robert P.",
    title = "{Spinor structure of space-times in general relativity. i}",
    doi = "10.1063/1.1664507",
    journal = "J. Math. Phys.",
    volume = "9",
    pages = "1739--1744",
    year = "1968"
}

@article{Geroch:1970uv,
    author = "Geroch, Robert P.",
    title = "{Spinor structure of space-times in general relativity. II}",
    doi = "10.1063/1.1665067",
    journal = "J. Math. Phys.",
    volume = "11",
    pages = "343--348",
    year = "1970"
}

@article{Geroch:1973am,
    author = "Geroch, Robert P. and Held, A. and Penrose, R.",
    title = "{A space-time calculus based on pairs of null directions}",
    doi = "10.1063/1.1666410",
    journal = "J. Math. Phys.",
    volume = "14",
    pages = "874--881",
    year = "1973"
}

@book{Penrose:1985bww,
    author = "Penrose, Roger and Rindler, Wolfgang",
    title = "{Spinors and Space-Time}",
    doi = "10.1017/CBO9780511564048",
    isbn = "978-0-521-33707-6, 978-0-511-86766-8, 978-0-521-33707-6",
    publisher = "Cambridge Univ. Press",
    address = "Cambridge, UK",
    series = "Cambridge Monographs on Mathematical Physics",
    month = "4",
    year = "2011"
}

@article{Newman:1961qr,
    author = "Newman, Ezra and Penrose, Roger",
    title = "{An Approach to gravitational radiation by a method of spin coefficients}",
    doi = "10.1063/1.1724257",
    journal = "J. Math. Phys.",
    volume = "3",
    pages = "566--578",
    year = "1962"
}

@article{Gourgoulhon:2005ng,
    author = "Gourgoulhon, Eric and Jaramillo, Jose Luis",
    title = "{A 3+1 perspective on null hypersurfaces and isolated horizons}",
    eprint = "gr-qc/0503113",
    archivePrefix = "arXiv",
    doi = "10.1016/j.physrep.2005.10.005",
    journal = "Phys. Rept.",
    volume = "423",
    pages = "159--294",
    year = "2006"
}

@article{Chandrasekaran:2021hxc,
    author = "Chandrasekaran, Venkatesa and Flanagan, Eanna E. and Shehzad, Ibrahim and Speranza, Antony J.",
    title = "{Brown-York charges at null boundaries}",
    eprint = "2109.11567",
    archivePrefix = "arXiv",
    primaryClass = "hep-th",
    doi = "10.1007/JHEP01(2022)029",
    journal = "JHEP",
    volume = "01",
    pages = "029",
    year = "2022"
}

@article{Saw:2016isu,
    author = "Saw, Vee-Liem",
    title = "{Mass-loss of an isolated gravitating system due to energy carried away by gravitational waves with a cosmological constant}",
    eprint = "1605.05151",
    archivePrefix = "arXiv",
    primaryClass = "gr-qc",
    doi = "10.1103/PhysRevD.94.104004",
    journal = "Phys. Rev. D",
    volume = "94",
    number = "10",
    pages = "104004",
    year = "2016"
}

@article{Mao:2019ahc,
    author = "Mao, Pujian",
    title = "{Asymptotics with a cosmological constant: The solution space}",
    eprint = "1901.04010",
    archivePrefix = "arXiv",
    primaryClass = "gr-qc",
    reportNumber = "CJQS-2019-015",
    doi = "10.1103/PhysRevD.99.104024",
    journal = "Phys. Rev. D",
    volume = "99",
    number = "10",
    pages = "104024",
    year = "2019"
}

@article{Compere:2019bua,
    author = "Comp{\`e}re, Geoffrey and Fiorucci, Adrien and Ruzziconi, Romain",
    title = "{The $\Lambda$-BMS$_4$ group of dS$_4$ and new boundary conditions for AdS$_4$}",
    eprint = "1905.00971",
    archivePrefix = "arXiv",
    primaryClass = "gr-qc",
    doi = "10.1088/1361-6382/ab3d4b",
    journal = "Class. Quant. Grav.",
    volume = "36",
    number = "19",
    pages = "195017",
    year = "2019",
    note = "[Erratum: Class.Quant.Grav. 38, 229501 (2021)]"
}

@article{Compere:2020lrt,
    author = "Comp{\`e}re, Geoffrey and Fiorucci, Adrien and Ruzziconi, Romain",
    title = "{The $\Lambda$-BMS$_4$ charge algebra}",
    eprint = "2004.10769",
    archivePrefix = "arXiv",
    primaryClass = "hep-th",
    doi = "10.1007/JHEP10(2020)205",
    journal = "JHEP",
    volume = "10",
    pages = "205",
    year = "2020"
}

@article{Taylor:2023ajd,
    author = "Taylor, Tomasz R. and Zhu, Bin",
    title = "{$w_{1+\infty}$ Algebra with a Cosmological Constant and the Celestial Sphere}",
    eprint = "2312.00876",
    archivePrefix = "arXiv",
    primaryClass = "hep-th",
    doi = "10.1103/PhysRevLett.132.221602",
    journal = "Phys. Rev. Lett.",
    volume = "132",
    number = "22",
    pages = "221602",
    year = "2024"
}

@article{Chandrasekaran:2018aop,
    author = "Chandrasekaran, Venkatesa and Flanagan, {\'E}anna {\'E}. and Prabhu, Kartik",
    title = "{Symmetries and charges of general relativity at null boundaries}",
    eprint = "1807.11499",
    archivePrefix = "arXiv",
    primaryClass = "hep-th",
    doi = "10.1007/JHEP11(2018)125",
    journal = "JHEP",
    volume = "11",
    pages = "125",
    year = "2018",
    note = "[Erratum: JHEP 07, 224 (2023)]"
}

\end{document}